\documentclass[aps,prd,onecolumn,preprintnumbers,notitlepage,superscriptaddress,
nofootinbib,amsfont,amssymb,10pt,singlespacing] {revtex4-1}
\usepackage{amsmath,bm}
\usepackage{times}
\usepackage{braket}
\usepackage{color,graphicx}
\usepackage{slashed}
\usepackage{hyperref}
\usepackage{graphics}
\usepackage{subfigure}
\usepackage{multirow,makecell}
\usepackage{textcomp}
\usepackage{mathtools}
\usepackage{float}

\newcommand{\beq}{\begin{eqnarray}}
\newcommand{\eeq}{\end{eqnarray}}

\def\lsim{ {\ \lower-1.2pt\vbox{\hbox{\rlap{$<$}\lower6pt\vbox{\hbox{$\sim$}
}}}\ } }
\def\gsim{ {\ \lower-1.2pt\vbox{\hbox{\rlap{$>$}\lower6pt\vbox{\hbox{$\sim$}
}}}\ } }
\definecolor{Red}{rgb}{1.,0.,0.}

\definecolor{Blue}{rgb}{0.,0.,1.}

\definecolor{nicered}{rgb}{0.7,0.1,0.1}
\definecolor{nicegreen}{rgb}{0.1,0.5,0.1}

\hypersetup{colorlinks,citecolor=nicegreen,linkcolor=nicered}

\begin{document}

\title{Contributions of S, P, and D-wave resonances to the quasi-two-body decays $B^{0}_{s} \rightarrow \psi(3686,3770)\emph{K}\pi$ in the perturbative QCD approach}

\author{Wen~Liu}
\email[Electronic address:]{625269181@qq.com}
\affiliation{School of Physical Science and Technology,
 Southwest University, Chongqing 400715, China}

\author{Xian-Qiao~Yu}
\email[Electronic address:]{yuxq@swu.edu.cn}
\affiliation{School of Physical Science and Technology,
Southwest University, Chongqing 400715, China}

\date{\today}

\begin{abstract}

Based on the perturbative quantum chromodynamics (pQCD) approach and the quasi-two-body approximation, we have studied the three-body decays $B^{0}_{s}\rightarrow \psi(3686,3770)\emph{K}\pi$, which include the contributions of the intermediate resonances $\overline{\emph{K}}^{*}_{0}(1430)^{0}$, $\overline{\emph{K}}^{*}(892)^{0}$, $\overline{\emph{K}}^{*}(1410)^{0}$, $\overline{\emph{K}}^{*}(1680)^{0}$, and $\overline{\emph{K}}^{*}_{2}(1430)^{0}$. The time-like form factors corresponding to the distribution amplitudes of the S, P, and D-wave of the kaon-pion pair have been adopted in the parameterized form, which describe the interactions between $\emph{K}$ and $\pi$ in the resonance region. First, the decays $B^{0}_{s}\rightarrow \psi(2S,1D)\emph{K}^{-}\pi^{+}$ have been calculated followed by the calculation of the branching ratios of the decays $B^{0}_{s}\rightarrow \psi(3686,3770)\emph{K}^{-}\pi^{+}$ using the 2S-1D mixing scheme. In addition, the pQCD predictions for the decays $B^{0}_{s}\rightarrow \psi(2S,1D)\emph{K}\pi$ and $B^{0}_{s}\rightarrow \psi(3686,3770)\emph{K}\pi$ have been obtained using the narrow-width approximation relation given by the Clebsch-Gorden coefficients. Our work shows that the $\overline{\emph{K}}^{*}(892)^{0}$ resonance is the main contributor to the total decay, and the branching ratio and the longitudinal polarization fraction of the $\psi(2S)\overline{\emph{K}}^{*}(892)^{0}$ decay mode agree well with the currently available data within errors. Furthermore, the theoretical predictions of the $\psi(2S)$ and $\psi(3686)$ decay modes are very close, indicating that they can be regarded as the same meson state. Finally, the pQCD predictions for branching ratios of decays $B^{0}_{s}\rightarrow \psi(3686,3770)\emph{K}\pi$ are of the order of $10^{-5}$ and $10^{-6}$, respectively, which can be verified using the ongoing LHCb and Belle II experiments.

\end{abstract}

\maketitle

%
%

\section{Introduction}\label{sec:intro}

In recent years, studies on B-meson decays have attracted increasing attention since they enable the testing of the standard model (SM) and enrich the field of quantum chromodynamics (QCD). The three-body decays of the B meson involve resonant as well as for non-resonant contributions. Thus their calculations are more complicated than those for two-body decays. There are mutual interferences between the resonant and non-resonant states, and thus it is difficult to calculate them separately~\cite{760940062007}. Based on the symmetry principles and the factorization theorems, a few theoretical models for calculating the three-body decay have been developed. In this study, we have adopted the widely used perturbative QCD (pQCD) factorization approach~\cite{980130042018,970340332018,771992017,890740312014}. The color-suppressed phenomenon occurs when a $B^{0}_{s}$ meson decays into a kaon-pion pair and a charmonium. Thus, it is meaningful to study the $B^{0}_{s}\rightarrow \psi(2S,1D)\emph{K}\pi$ decays. Recently, significant advances have been made in the research on heavy quarkonium generation mechanism~\cite{753112015}. The LHCb collaboration has detected the $B^{0}_{s}\rightarrow \psi(2S)\emph{K}^{-}\pi^{+}$ decay~\cite{7474842015} and found that the main source of the decay branching ratio is the $\overline{\emph{K}}^{*}(892)^{0}$ resonance. These advances have allowed us to reliably calculate and test the $B^{0}_{s}\rightarrow \psi(2S,1D)\emph{K}\pi$ decays.

The pQCD factorization approach was proposed based on the $\emph{k}_{T}$ factorization theorem~\cite{3811291982,5612582003,700540062004}. In this approach, a three-body problem can be simplified to a quasi-two-body problem by introducing two-hadron distribution amplitudes(DAs)~\cite{8117821998,620730142000}. The predominant contributions in the decay process are from the parallel motion range, where the invariant mass of the double light meson pair is lower than $\emph{O}(\bar{\Lambda}M_{B})$, and $\bar{\Lambda}=M_{B}-m_{b}$ represents the mass difference between the $B$ meson and the b-quark. Thus, the pQCD factorization formula for the three-body decay of the $B^{0}_{s}$ meson can be generally described as~\cite{5612582003,700540062004}
\begin{equation}
{\cal A}={\cal H}\otimes \phi_{\emph{B}^{0}_{s}} \otimes \phi_{h_{3}} \otimes \phi_{h_{1}h_{2}},
\label{eq:exp1}
\end{equation}
where the hard decay kernel, ${\cal H}$, represents the contribution of the Feynman diagram with only one gluon exchange in the leading order, which can be calculated using the perturbation theory. The terms $\phi_{\emph{B}^{0}_{s}}$, $\phi_{h_{3}}$, and $\phi_{h_{1}h_{2}}$ represent the wave functions of $B^{0}_{s}$, $h_3$, and $h_1h_2$ pair, respectively. They are considered as non-perturbative inputs, which can be constructed by extracting the relevant experimentally measured quantities or calculating them using the non-perturbative model.

Though the decay $B^{0}_{s}\rightarrow \psi(3770)\emph{K}\pi$ has not been observed experimentally, the mixing structure of $\psi(3770)$ can be investigated by making a theoretical prediction for this decay channel. Since the charmonium mesons $\psi(3686)$ and $\psi(3770)$ are regarded as the 2S-1D mixed states, the decays $B^{0}_{s}\rightarrow \psi(2S)\emph{K}\pi$ and $B^{0}_{s}\rightarrow \psi(1D)\emph{K}\pi$ should first be calculated, and then the fitting should be performed based on the 2S-1D mixing scheme to obtain the branching ratios of the decays $B^{0}_{s}\rightarrow \psi(3686,3770)\emph{K}\pi$. The $\psi(1D)$ state denotes the orbital quantum number $l=2$ and the principal quantum number $n=1$, and $\psi(2S)$ is the first radially excited state of the charmonium meson.

The 2S-1D mixing angle, $\theta$, is related to the ratio of the lepton decay widths of $\psi(3686)$ and $\psi(3770)$~\cite{59212004}, and its value can be obtained from the fitting of the non-relativistic potential model~\cite{640940022001,650940242002,4435621991}. The theoretical prediction for the $B\rightarrow \psi(3770)\emph{K}$ decay is in line with the experimental measurement when higher-twist effects are considered and the 2S-1D mixing angle of $\theta=-(12\pm2)^{\circ}$ has been adopted~\cite{283612006}. In addition, two mixing angle options, namely, $\theta=(27\pm2)^{\circ}$ and $\theta=-(12\pm2)^{\circ}$, have been offered~\cite{640940022001,650940242002,4435621991}. Based on these views, $\psi(3686)$ and $\psi(3770)$ can be represented as follows~\cite{4435621991,970960082018}:
\begin{equation}
\begin{split}
\psi(3686)=\sin\theta|c\bar{c}(1D)\rangle+\cos\theta|c\bar{c}(2S)\rangle,   \\
\psi(3770)=\cos\theta|c\bar{c}(1D)\rangle-\sin\theta|c\bar{c}(2S)\rangle.
\end{split}
\end{equation}

The branching ratio is affected by the width of the resonant state and the interactions between the final-state meson pair, especially the direct CP violations. Hence, introducing an intermediate resonance, $\overline{\emph{K}}^{*0}$, is more appropriate~\cite{763292016,766752016,950560082017}. We consider the contributions of the S, P, and D-wave resonances from the kaon-pion pair in the quasi-two-body decays $B^{0}_{s}\rightarrow \psi(2S,1D)(\overline{\emph{K}}^{*0}\rightarrow)\emph{K}\pi$. In this work, the contributions of the following five intermediate resonances have been included: $\overline{\emph{K}}^{*}_{0}(1430)^{0}$, $\overline{\emph{K}}^{*}(892)^{0}$, $\overline{\emph{K}}^{*}(1410)^{0}$, $\overline{\emph{K}}^{*}(1680)^{0}$, and $\overline{\emph{K}}^{*}_{2}(1430)^{0}$. According to Eq.~(\ref{eq:exp1}), $\phi_{h_{3}}$ denotes the wave functions of the charmonium $\psi$ and $\phi_{h_{1}h_{2}}$ represent the various partial-wave functions of the kaon-pion pair, such as S-wave $\overline{\emph{K}}^{*}_{0}(1430)^{0}$, P-wave $\overline{\emph{K}}^{*}(892)^{0}$, and D-wave $\overline{\emph{K}}^{*}_{2}(1430)^{0}$. We refer to the study by Rui and Wang~\cite{970330062018} to obtain the information of the S-wave DAs. For the P-wave, there are three possible polarizations: longitudinal, parallel and perpendicular amplitudes. Hence, we have considered both the longitudinal as well as transverse polarization cases of the P-wave DAs. The P-wave DAs have been described analogously to the two-pion DAs~\cite{981130032018}, which include the longitudinal polarization fraction and the flavor-symmetry-breaking effect. At present, studies on the D-wave DAs are inadequate, and thus we have adopted the method used in the study by Rui et al.~\cite{797922019} to construct the D-wave DAs using a similar method of the $\emph{K}\emph{K}$ pair.

The contents of this paper have been organized as follows. In Section. \ref{sec:pert}, a description has been given of the computational framework and a list of the wave functions involved in this work. Expressions for the various decay amplitudes associated with the theoretical calculations have been presented in Section. \ref{sec:amp}. Section. \ref{sec:numer} presents the numerical results and the related discussions. The study has been summarized in Section. \ref{sec:summary}.

%
%
\section{Computational framework}\label{sec:pert}

The weak-effective Hamiltonian of the $B^{0}_{s}\rightarrow \psi(2S,1D)\overline{\emph{K}}^{*0}(\rightarrow\emph{K}^{-}\pi^{+})$ decays is expressed as~\cite{6811251996}
\begin{equation}
{\cal H}_{eff}=\frac{G_{F}}{\sqrt2}\big\{V^*_{cb}V_{cd}[C_{1}O_{1}+C_{2}O_{2}]-V^*_{tb}V_{td}[\sum^{10}_{i=3}C_{i}O_{i}]\big\},
\end{equation}
where $V^*_{cb}$$V_{cd}$ and $V^{*}_{tb}$$V_{td}$ are the CKM factors, $O_{i}$ is the localized four-quark operator, and $C_{i}$ is the Wilson coefficient corresponding to the quark operator.

 \begin{figure}[htbp]
 \centering
 \begin{tabular}{l}
 \includegraphics[width=0.5\textwidth]{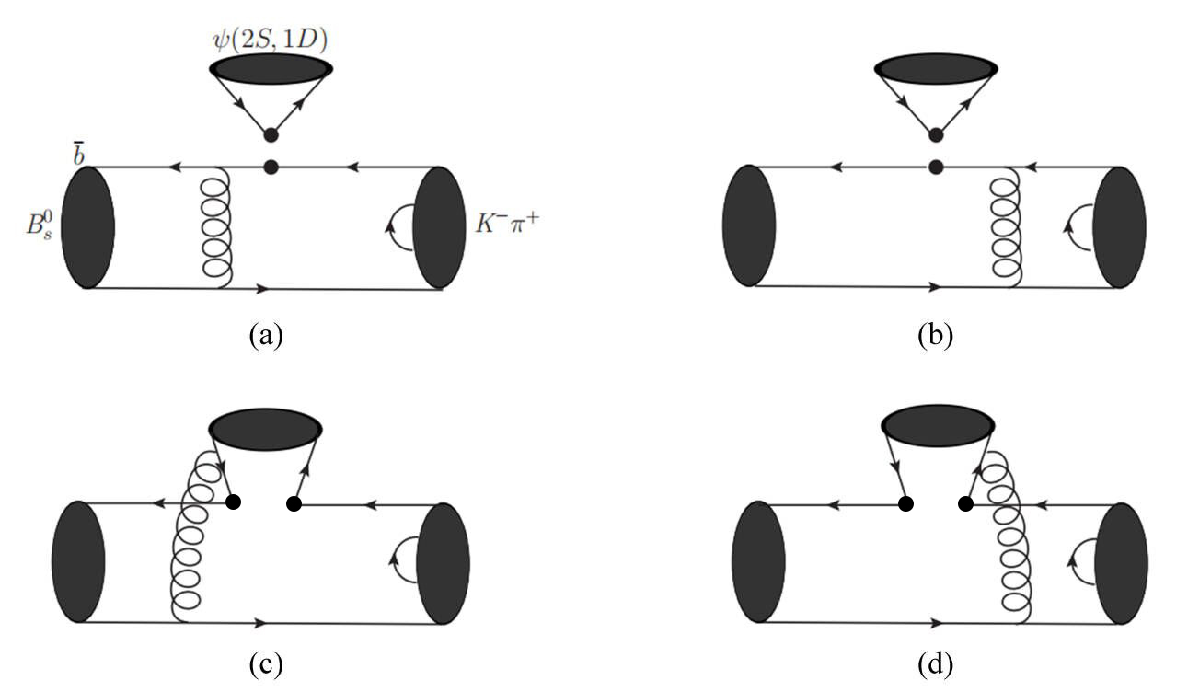}
 \end{tabular}
 \caption {Feynman diagrams of the $B^{0}_{s}\rightarrow \psi(2S,1D)(\overline{\emph{K}}^{*0}\rightarrow)\emph{K}^{-}\pi^{+}$ decays}
 \label{Feynman}
 \end{figure}

To simplify the calculation, we have chosen to describe the decay process in the light-cone coordinate system. Assuming that the initial state of the $B^{0}_{s}$ meson is stationary, the charmonium $\psi(2S,1D)$ and the $\emph{K}\pi$ pair move in the directions of the vectors $v=(0,1,0_{T})$ and $n=(1,0,0_{T})$, respectively. The Feynman diagrams of the decay are depicted in Fig.~\ref{Feynman}. $\emph{p}_{B}$, $\emph{p}$ and $\emph{p}_{3}$ represent the momenta of the $B^0_{s}$, $\overline{\emph{K}}^{*0}$ and $\psi(2S,1D)$ mesons, respectively.
\begin{equation}
\begin{split}
&\emph{p}_{B}=\frac{M_{B^{0}_{s}}}{\sqrt{2}}(1,1,0_{T}),                 \\
&\emph{p}=\frac{M_{B^{0}_{s}}}{\sqrt{2}}(1-r^2,\eta,0_{T}),              \\
&\emph{p}_{3}=\frac{M_{B^{0}_{s}}}{\sqrt{2}}(r^2,1-\eta,0_{T}).
\end{split}
\end{equation}

In addition, the momenta of the light quark corresponding to the $B^0_{s}$, $\overline{\emph{K}}^{*0}$ and $\psi(2S,1D)$ mesons, respectively, are as follows:
\begin{equation}
\begin{split}
&\emph{k}_{B}=(0,\frac{M_{B^{0}_{s}}}{\sqrt{2}}x_{B},\emph{k}_{BT}),\\
&\emph{k}=(\frac{M_{B^{0}_{s}}}{\sqrt{2}}z(1-r^{2}),0,\emph{k}_{T}),\\
&\emph{k}_{3}=(\frac{M_{B^{0}_{s}}}{\sqrt{2}}r^{2}x_{3},\frac{M_{B^{0}_{s}}}{\sqrt{2}}(1-\eta)x_{3},\emph{k}_{3T}),
\end{split}
\end{equation}
where $M_{B^{0}_{s}}$ represents the mass of the $B^{0}_{s}$ meson, $M_{\psi}$ is the mass of the charmonium $\psi(2S,1D)$, $r=\frac{M_\psi}{M_{B^{0}_{s}}}$, the variable $\eta=\omega^2/({M_{B^{0}_{s}}^2}-{M_\psi}^2)$, and $\omega$ represents the invariant mass of the kaon-pion pair, which conforms to the relationship $\omega^2=\emph{p}^2$. $\emph{x}_{B}$, $\emph{z}$, and $\emph{x}_{3}$ are the proportions of the momenta of the spectator quark inside the $B^{0}_{s}$, $\overline{\emph{K}}^{*0}$, and $\psi(2S,1D)$ mesons, respectively, with values in the range of $[0,1]$.

Then, the momenta $\emph{p}_{1}$ and $\emph{p}_{2}$ are defined in the kaon-pion pair as follows:
\begin{equation}
\begin{split}
\emph{p}_{1}=(\zeta \emph{p}^{+},\eta(1-\zeta) \emph{p}^{+},\emph{p}_{1T}),\\
\emph{p}_{2}=((1-\zeta)\emph{p}^{+},\eta\zeta \emph{p}^{+},\emph{p}_{2T}).
\end{split}
\end{equation}
The variable $\zeta=\frac{p^{+}_{1}}{p^{+}}$ depicts the distribution of the longitudinal momentum of the kaon with $\emph{p}^{2}_{1T}=\emph{p}^{2}_{2T}=(1-\zeta)\zeta\omega^{2}$.

The $B^{0}_{s}$ meson is considered a heavy-light model, and its wave function is expressed as~\cite{630740092001,630540082001,50462001}
\begin{equation}
\Phi_{B^{0}_{s}}=\frac{\emph{i}}{\sqrt{2N_{c}}}(\not {p}_{B}+M_{B^{0}_{s}}){\gamma_{5}}{\phi_{B_{s}}({x_{B},b_{B}})}.
\end{equation}
The distribution amplitude(DA) ${\phi_{B_{s}}({x_{B},b_{B}})}$ is expressed as
\begin{equation}
\phi_{B_{s}}(x_{B},b_{B})=\emph{N}_{B}{{x_{B}}^2}(1-{x_{B}})^2\times\exp\bigg[-\frac{M^2_{B^{0}_{s}}{{x_{B}}^2}}{2\omega^2_{B_{s}}}-\frac{1}{2}(\omega_{B_{s}}{b_{B}})^2\bigg],
\end{equation}
where $\emph{N}_{B}$ is the normalization factor, and its value can be obtained using the normalization relation $\int^{1}_{0}\emph{dx}_{B}\phi_{B_{s}}(x_{B},b_{B}=0)=\emph{f}_{B^{0}_{s}}/({2}{\sqrt{{2}{\emph{N}}_{c}}})$. Here, the color factor $N_c=3$, and we select the shape parameter $\omega_{B_{s}}=0.50\pm0.05$ {\rm GeV}~\cite{760740182007} have been used.

We have applied the wave function form described on the basis of the harmonic oscillator for the $\psi(3686)$ and $\psi(3770)$ states. This form has been successfully applied to many charmonium mesons, such as $\psi(2S)$, $\psi(3S)$, and $J/\psi$~\cite{970960082018,810375012010,752932015,765642016}. The longitudinally and transversely polarized wave functions of $\psi(2S)$ and $\psi(1D)$ are expressed as follows~\cite{981130032018,752932015,765642016}:
\begin{equation}
\begin{split}
&\Phi^{L}_{\psi}=\frac{1}{\sqrt{2\emph{N}_{c}}}[M_{\psi}\not{\epsilon}_{3L}\psi^L(x_{3},b_{3})+\not{\epsilon}_{3L}\not {p}_{3}\psi^t(x_{3},b_{3})],\\ &\Phi^{T}_{\psi}=\frac{1}{\sqrt{2\emph{N}_{c}}}[M_{\psi}\not{\epsilon}_{3T}\psi^V(x_{3},b_{3})+\not{\epsilon}_{3T}\not {p}_{3}\psi^T(x_{3},b_{3})],
\end{split}
\end{equation}
where $\emph{p}_{3}$ represents the momentum of the $\psi(2S,1D)$ meson and $M_{\psi}$ is its mass. The longitudinal polarization vector $\epsilon_{3L}=\frac{M_{B^0_{s}}}{\sqrt{2} M_{\psi}}(-{r^2},(1-\eta),0_{T})$ and the transverse polarization vector $\epsilon_{3T}=(0,0,1_{T})$. The twist-2 and twist-3 DAs are as follows~\cite{970960082018,752932015}:
\begin{equation}
\psi^{\emph{L,T}}(\emph{x}_{3},\emph{b}_{3})= \frac {\emph{f}_{(2S,1D)}}{2\sqrt{2N_{c}}}N^{\emph{L,T}}\emph{x}_{3}\overline{\emph{x}}_{3} {\cal I}(\emph{x}_{3})\times \exp\bigg[-\emph{x}_{3}\overline{\emph{x}}_{3}\frac{\emph{m}_c}{w}
\bigg[w^2\emph{b}^2_{3}+\bigg(\frac{\emph{x}_{3}-\overline{\emph{x}}_{3}}{2\emph{x}_{3}\overline{\emph{x}}_{3}}\bigg)^2\bigg]\bigg],
\label{eq:exp10}
\end{equation}

\begin{equation}
\psi^{\emph{t}}(\emph{x}_{3},\emph{b}_{3})= \frac {\emph{f}_{(2S,1D)}}{2\sqrt{2N_{c}}}N^{\emph{t}} ({\emph{x}_{3}-\overline{\emph{x}}_{3}})^2{\cal I}(\emph{x}_{3})\times\exp\bigg[-\emph{x}_{3}\overline{\emph{x}}_{3}\frac{\emph{m}_c}{w}
\bigg[w^2\emph{b}^2_{3}+\bigg(\frac{\emph{x}_{3}-\overline{\emph{x}}_{3}}{2\emph{x}_{3}\overline{\emph{x}}_{3}}\bigg)^2\bigg]\bigg],
\label{eq:exp11}
\end{equation}

\begin{equation}
\psi^{\emph{V}}(\emph{x}_{3},\emph{b}_{3})= \frac {\emph{f}_{(2S,1D)}}{2\sqrt{2N_{c}}}N^{\emph{V}} [1+({\emph{x}_{3}-\overline{\emph{x}}_{3}})^2]{\cal I}(\emph{x}_{3})\times\exp\bigg[-\emph{x}_{3}\overline{\emph{x}}_{3}\frac{\emph{m}_c}{w}
\bigg[w^2\emph{b}^2_{3}+\bigg(\frac{\emph{x}_{3}-\overline{\emph{x}}_{3}}{2\emph{x}_{3}\overline{\emph{x}}_{3}}\bigg)^2\bigg]\bigg],
\label{eq:exp12}
\end{equation}
where ${\cal I}(\emph{x}_{3})=(\frac{1}{\emph{x}_{3}\overline{\emph{x}}_{3}}-\emph{m}_{c}w \emph{b}^2_{3})(6\emph{x}^4_{3}-12\emph{x}^3_{3}+7\emph{x}^2_{3}-\emph{x}_{3})-\frac{\emph{m}_c(1-2\emph{x}_{3})^2}{4w \emph{x}_{3}\overline{\emph{x}}_{3}}$ for $\psi(1D)$ and ${\cal I}(\emph{x}_{3})=1-4{\emph{m}_c}w {\emph{x}_{3}\overline{\emph{x}}_{3}}\emph{b}^2_{3}+\frac{\emph{m}_c(1-2\emph{x}_{3})^2}{w \emph{x}_{3}\overline{\emph{x}}_{3}}$ for $\psi(2S)$. We have selected the shape parameters $w_{1D}=0.5\pm0.05$ {\rm GeV}~\cite{970960082018} and $w_{2S}=0.2\pm0.1$ {\rm GeV}~\cite{752932015}. The normalization factor $N^{\emph{i}}(\emph{i}=\emph{L,T,t,V})$ can be obtained using the normalization relationship $\int^1_0 \psi^{\emph{i}}(\emph{x}_{3},b_{3}=0)d\emph{x}_{3}= \frac {\emph{f}_{(2S,1D)}}{2\sqrt{2N_{c}}}$. Eqs.~(\ref{eq:exp10})$-$~(\ref{eq:exp12}) are symmetric under the transformation of $\emph{x}_{3}\leftrightarrow \overline{\emph{x}}_{3}$.

A form similar to the two-pion DA has been adopted for the S-wave of the kaon-pion pair DA~\cite{910940242015}:
\begin{equation}
\Phi_{S}=\frac{1}{\sqrt{2 N_{c}}}[\not {p}\phi^{0}_{S}(z,\zeta,\omega^2)+\omega\phi^{s}_{S}(z,\zeta,\omega^2)+\omega(\not{n}\not {v}-1)\phi^{t}_{S}(z,\zeta,\omega^2)].
\end{equation}
The subscripts S, P, and D denote the corresponding sub-waves, respectively, in the following description.

Using the description given by Wang et al.~\cite{1010160152020}, the twist-2 DAs have been described in a form similar to the scalar meson~\cite{730140172006,7303362014}, whereas asymptotic forms for the twist-3 DAs have been adopted in this work. They can be expressed as follows:
\begin{equation}
\phi^{0}_{S}(z,\zeta,\omega^2)=\frac{6}{2\sqrt{2 N_{c}}}\emph{F}_{\emph{S}}(\omega^2)(z-z^{2})\bigg[\frac{1}{\mu_{\emph{S}}}+\emph{B}_{1}\emph{C}^{3/2}_{1}(t)+\emph{B}_{3}\emph{C}^{3/2}_{3}(t)\bigg],
\end{equation}

\begin{equation}
\phi^{s}_{S}(z,\zeta,\omega^2)=\frac{1}{2\sqrt{2 N_{c}}}\emph{F}_{\emph{S}}(\omega^2),
\end{equation}

\begin{equation}
\phi^{t}_{S}(z,\zeta,\omega^2)=\frac{1}{2\sqrt{2 N_{c}}}\emph{F}_{\emph{S}}(\omega^2)(1-2z).
\end{equation}
The Gegenbauer polynomials are $\emph{C}^{3/2}_{1}(t)=3t$ and $\emph{C}^{3/2}_{3}(t)=\frac{5}{2}(7t^{3}-3t)$ with $t=1-2z$. In addition, $\mu_{\emph{S}}=\frac{\omega}{m_{2}-m_{1}}$, $\emph{m}_{1}$ and $\emph{m}_{2}$ represent the corresponding current quark masses, and the Gegenbauer moments are $\emph{B}_{1}=-0.57\pm0.13$ and $\emph{B}_{3}=-0.42\pm0.22$~\cite{730140172006,770140342008,780140062008}.

For the time-like scalar form factor, $\emph{F}_{\emph{S}}(\omega^2)$, we have adopted the parameterized fitting results of an improved LASS line type presented by Aston et al.~\cite{2964931988}. $\emph{F}_{\emph{S}}(\omega^2)$ is expressed as~\cite{1010160152020}
\begin{equation}
\emph{F}_{\emph{S}}(\omega^2)=\frac{m^{2}_{0}\frac{\Gamma_{0}}{\mid{\overrightarrow{p_{0}}}\mid}}{m^{2}_{0}-\omega^2-\emph{i}m^{2}_{0}\frac{\Gamma_{0}}{\omega}\frac{\mid{\overrightarrow{p_{1}}}\mid}{\mid{\overrightarrow{p_{0}}}\mid}}\emph{e}^{2\emph{i}\delta_{\emph{B}}}+\frac{\omega}{\mid{\overrightarrow{p_{1}}}\mid [\textmd{cot}(\delta_{\emph{B}})-\emph{i}]},
\label{eq:exp16}
\end{equation}

\begin{equation}
\textmd{cot}(\delta_{\emph{B}})=\frac{a{\mid{\overrightarrow{p_{1}}}\mid}}{2}+\frac{1}{l{\mid{\overrightarrow{p_{1}}}\mid}}.
\end{equation}
In Eq.~(\ref{eq:exp16}), the first term contains the resonant contribution with a phase factor to maintain unitarity, and the second term is an empirical term of the elastic $\emph{K}\pi$ scattering. $\Gamma_{0}$ and $m_{0}$ represent the width and the pole mass, respectively, of the $\overline{\emph{K}}^{*}_{0}(1430)^{0}$ resonance, $\mid{\overrightarrow{p_{1}}}\mid$ represents the momentum of the decay product of the intermediate resonance, and $\mid{\overrightarrow{p_{0}}}\mid=\mid{\overrightarrow{p_{1}}}\mid$ is available when $\omega=m_{\emph{K}^{*0}}$. $a=(7.0\pm2.4)$ ${\rm GeV^{-1}}$ and $l=(3.1\pm1.0)$ ${\rm GeV^{-1}}$ are the effective range and the scattering length, respectively, which are universal in describing the $\emph{K}\pi$ meson pair.

According to the Li et al.~\cite{440731022020}, the P-wave kaon-pion DAs related to the longitudinal and transverse polarizations can be expressed as
\begin{equation}
\begin{split}
&\Phi^{L}_{P}=\frac{1}{\sqrt{2 N_{c}}}\bigg[\not {p}\phi^{0}_{P}(z,\zeta,\omega^2)+\omega\phi^{s}_{P}(z,\zeta,\omega^2)+\frac{\not{p}_{1}\not{p}_{2}-\not{p}_{2}\not{p}_{1}}{\omega(2\zeta-1)}\phi^{t}_{P}(z,\zeta,\omega^2)\bigg],\\
&\Phi^{T}_{P}=\frac{1}{\sqrt{2 N_{c}}}\bigg[\gamma_{5}\not{\epsilon}_{T}\not {p}\phi^{T}_{P}(z,\zeta,\omega^2)+\omega \gamma_{5} \not{\epsilon}_{T} \phi^{a}_{P}(z,\zeta,\omega^2)+\emph{i} \omega \frac{\epsilon^{\mu\nu\rho\sigma}\gamma_{\mu}\epsilon_{T\nu}P_{\rho}n_{-\sigma}}{P\cdot n_{-}}\phi^{v}_{P}(z,\zeta,\omega^2)\bigg].
\label{eq:exp18}
\end{split}
\end{equation}

The different twists in Eq.~(\ref{eq:exp18}) when expanded using the Gegenbauer polynomial have the specific forms as follows:
\begin{equation}
\phi^{0}_{P}(z,\zeta,\omega^2)=\frac{3}{\sqrt{2 N_{c}}}\emph{F}^{\|}_{\emph{P}}(\omega^2)(z-z^{2})\bigg[1+3ta^{\|}_{1\emph{K}^{*}}+\frac{3}{2}(5t^{2}-1)a^{\|}_{2\emph{K}^{*}}\bigg](2\zeta-\alpha-1),
\end{equation}

\begin{equation}
\phi^{s}_{P}(z,\zeta,\omega^2)=\frac{3}{2\sqrt{2 N_{c}}}\emph{F}^{\bot}_{\emph{P}}(\omega^2)[t(1+ta^{\bot}_{1s})-(2z-2z^{2})a^{\bot}_{1s}](2\zeta-1),
\end{equation}

\begin{equation}
\phi^{t}_{P}(z,\zeta,\omega^2)=\frac{3}{2\sqrt{2 N_{c}}}\emph{F}^{\bot}_{\emph{P}}(\omega^2)[t^{2}+(3t^{3}-t)a^{\bot}_{1t}](2\zeta-1),
\end{equation}

\begin{equation}
\phi^{T}_{P}(z,\zeta,\omega^2)=\frac{3}{\sqrt{2 N_{c}}}\emph{F}^{\bot}_{\emph{P}}(\omega^2)(z-z^{2})\bigg[1+3ta^{\bot}_{1\emph{K}^{*}}+\frac{3}{2}(5t^{2}-1)a^{\bot}_{2\emph{K}^{*}}\bigg]\sqrt{\zeta-\zeta^{2}},
\end{equation}

\begin{equation}
\phi^{a}_{P}(z,\zeta,\omega^2)=\frac{3}{4\sqrt{2N_{c}}}\emph{F}^{\|}_{\emph{P}}(\omega^2) [t(1+ta^{\|}_{1a})-(2z-2z^{2})a^{\|}_{1a}]\sqrt{\zeta-\zeta^{2}},
\end{equation}

\begin{equation}
\phi^{v}_{P}(z,\zeta,\omega^2)=\frac{3}{8\sqrt{2 N_{c}}}\emph{F}^{\|}_{\emph{P}}(\omega^2)[1+t^{2}+t^{3}a^{\|}_{1v}]\sqrt{\zeta-\zeta^{2}}.
\end{equation}
The $SU(3)$ asymmetry factor $\alpha=(m^{2}_{\emph{K}^{\pm}}-m^{2}_{\pi^{\pm}})/\omega^2$, and the Gegenbauer moments $a^{\|}_{1\emph{K}^{*}}=0.2\pm0.2$, $a^{\|}_{2\emph{K}^{*}}=0.5\pm0.5$, $a^{\bot}_{1\emph{K}^{*}}=0.3\pm0.3$, $a^{\bot}_{2\emph{K}^{*}}=0.8\pm0.8$, $a^{\bot}_{1s}=-0.2$, $a^{\bot}_{1t}=0.2$, $a^{\|}_{1a}=-0.3$ and $a^{\|}_{1v}=0.3$~\cite{440731022020} have been adopted in this work.

The time-like shape factor, $\emph{F}^{\|}_{\emph{P}}(\omega^2)$, of the P-wave is expressed as~\cite{7810192018}
\begin{equation}
\emph{F}^{\|}_{\emph{P}}(\omega^2)=\frac{c_{1}m^{2}_{\emph{K}^{*}(892)^{0}}}{m^{2}_{\emph{K}^{*}(892)^{0}}-\omega^2-\emph{i}m_{\emph{K}^{*}(892)^{0}}\Gamma_{1}(\omega^2)}+\frac{c_{2}m^{2}_{\emph{K}^{*}(1410)^{0}}}{m^{2}_{\emph{K}^{*}(1410)^{0}}-\omega^2-\emph{i}m_{\emph{K}^{*}(1410)^{0}}\Gamma_{2}(\omega^2)}
+\frac{c_{3}m^{2}_{\emph{K}^{*}(1680)^{0}}}{m^{2}_{\emph{K}^{*}(1680)^{0}}-\omega^2-\emph{i}m_{\emph{K}^{*}(1680)^{0}}\Gamma_{3}(\omega^2)}.
\label{eq:exp22}
\end{equation}
The three terms added together have been derived from the $\emph{K}^{*}(892)^{0}$, $\emph{K}^{*}(1410)^{0}$, and $\emph{K}^{*}(1680)^{0}$ resonant states, and their corresponding weight coefficients are $c_{1}=0.72$, $c_{2}=0.134$, and $c_{3}=0.143$~\cite{440731022020}, respectively.

The mass-related width is given by
\begin{equation}
\Gamma_{\emph{i}}(\omega^2)=\Gamma_{\emph{i}}\bigg(\frac{m_{\emph{i}}}{\omega}\bigg)\bigg(\frac{\mid{\overrightarrow{p_{1}}}\mid}{\mid{\overrightarrow{p_{0}}}\mid}\bigg)^{(2\emph{L}_{\emph{R}}+1)},
\end{equation}
where $\Gamma_{\emph{i}}$ and $m_{\emph{i}}$ denote the width and the pole mass, respectively, of the corresponding resonance, $\emph{L}_{\emph{R}}$ represents the orbital angular momentum, with values of 0, 1 and 2 for the S, P, and D-wave, respectively. According to the study by Wang and Li~\cite{763292016}, the following relation can be obtained
\begin{equation}
\frac{\emph{F}^{\bot}_{\emph{P}}(\omega^2)}{\emph{F}^{\|}_{\emph{P}}(\omega^2)}\approx\frac{\emph{f}^{\emph{T}}_{\emph{K}^{*}}}{\emph{f}_{\emph{K}^{*}}},
\end{equation}
where $\emph{f}^{\emph{T}}_{\emph{K}^{*}}=0.185\pm0.010$ {\rm GeV} and $\emph{f}_{\emph{K}^{*}}=0.217\pm0.005$ {\rm GeV}~\cite{760740182007}. We have adopted the procedure from the work by Li et al.~\cite{440731022020}: studies on the decay constants of $\emph{K}^{*}(1410)^{0}$ and $\emph{K}^{*}(1680)^{0}$ are limited, and thus we have used the two decay constants of $\emph{K}^{*}(892)^{0}$ to determine the ratio $\emph{f}^{\emph{T}}_{\emph{K}^{*}}/\emph{f}_{\emph{K}^{*}}$.

A form similar to the two-kaon DAs has also been considered in the D-wave kaon-pion DAs~\cite{797922019}:
\begin{equation}
\begin{split}
&\Phi^{L}_{D}=\sqrt{\frac{2}{3}}\frac{1}{\sqrt{2 N_{c}}}\bigg[\not {p}\phi^{0}_{D}(z,\zeta,\omega^2)+\omega\phi^{s}_{D}(z,\zeta,\omega^2)+\frac{\not{p}_{1}\not{p}_{2}-\not{p}_{2}\not{p}_{1}}{\omega(2\zeta-1)}\phi^{t}_{D}(z,\zeta,\omega^2)\bigg],\\
&\Phi^{T}_{D}=\sqrt{\frac{1}{2}}\frac{1}{\sqrt{2 N_{c}}}\bigg[\gamma_{5}\not{\epsilon}_{T}\not {p}\phi^{T}_{D}(z,\zeta,\omega^2)+\omega \gamma_{5} \not{\epsilon}_{T} \phi^{a}_{D}(z,\zeta,\omega^2)+\emph{i} \omega \frac{\epsilon^{\mu\nu\rho\sigma}\gamma_{\mu}\epsilon_{T\nu}P_{\rho}n_{-\sigma}}{P\cdot n_{-}}\phi^{v}_{D}(z,\zeta,\omega^2)\bigg],
\end{split}
\end{equation}
where the coefficient $\sqrt{\frac{2}{3}}(\sqrt{\frac{1}{2}})$ comes from the different definitions of the polarization vector between the vector and tensor mesons in the longitudinal(transverse) polarization.

The different twists in the D-wave DAs are~\cite{797922019,830340012011,830140082011,860940152012}
\begin{equation}
\phi^{0}_{D}(z,\zeta,\omega^2)=\frac{9}{\sqrt{2 N_{c}}}\emph{F}^{\|}_{\emph{D}}(\omega^2)(z-z^{2})(2z-1)a^{0}_{1}(1-6\zeta+6\zeta^{2}),
\end{equation}

\begin{equation}
\phi^{s}_{D}(z,\zeta,\omega^2)=-\frac{9}{4\sqrt{2 N_{c}}}\emph{F}^{\bot}_{\emph{D}}(\omega^2)(1-6z+6z^{2})a^{0}_{1}(1-6\zeta+6\zeta^{2}),
\end{equation}

\begin{equation}
\phi^{t}_{D}(z,\zeta,\omega^2)=\frac{9}{4\sqrt{2 N_{c}}}\emph{F}^{\bot}_{\emph{D}}(\omega^2)(2z-1)(1-6z+6z^{2})a^{0}_{1}(1-6\zeta+6\zeta^{2}),
\end{equation}

\begin{equation}
\phi^{T}_{D}(z,\zeta,\omega^2)=\frac{9}{\sqrt{2 N_{c}}}\emph{F}^{\bot}_{\emph{D}}(\omega^2)(z-z^{2})(2z-1)a^{T}_{1}(2\zeta-1)\sqrt{\zeta-\zeta ^{2}},
\end{equation}

\begin{equation}
\phi^{a}_{D}(z,\zeta,\omega^2)=\frac{3}{2\sqrt{2 N_{c}}}\emph{F}^{\|}_{\emph{D}}(\omega^2)(2z-1)^{3}a^{T}_{1}(2\zeta-1)\sqrt{\zeta-\zeta ^{2}},
\end{equation}

\begin{equation}
\phi^{v}_{D}(z,\zeta,\omega^2)=-\frac{3}{2\sqrt{2 N_{c}}}\emph{F}^{\|}_{\emph{D}}(\omega^2)(1-6z+6z^{2})a^{T}_{1}(2\zeta-1)\sqrt{\zeta-\zeta ^{2}}.
\end{equation}
The Gegenbauer moments are $a^{0}_{1}=0.4\pm0.1$ and $a^{T}_{1}=0.8\pm0.2$, and a form similar to Eq.~(\ref{eq:exp22}) has been adopted for the time-like shape factor, $\emph{F}^{\|}_{\emph{D}}(\omega^2)$. Furthermore, the approximate relation $\emph{F}^{\bot}_{\emph{D}}(\omega^2)/\emph{F}^{\|}_{\emph{D}}(\omega^2)\approx\emph{f}^{\emph{T}}_{\emph{K}^{*}_{2}(1430)}/\emph{f}_{\emph{K}^{*}_{2}(1430)}$ can also be found, with $\emph{f}^{\emph{T}}_{\emph{K}^{*}_{2}(1430)}=0.077\pm0.014$ {\rm GeV} and $\emph{f}_{\emph{K}^{*}_{2}(1430)}=0.118\pm0.005$ {\rm GeV}~\cite{830340012011}.

The differential decay ratios for the $B^{0}_{s} \rightarrow \psi(2S,1D)\emph{K}^-\pi^+$ decays in the $B^{0}_{s}$ meson rest frame can be written as
\begin{equation}
\frac{d\cal{B}}{d\omega}=\frac{\tau_{B^0_{s}}\omega \mid {\overrightarrow{p_{1}}} \mid \mid {\overrightarrow{p_{3}}} \mid}{32(\pi M_{B^0_{s}})^3} \sum_{i=0,\|,\bot}\mid {\cal A}_{i} \mid^2,
\end{equation}
where the three-momenta of $\emph{K}^{-}$ and $\psi(2S,1D)$ in the kaon-pion center-of-mass system are expressed as
\begin{equation}
\begin{split}
&\mid {\overrightarrow{p_{1}}} \mid=\frac{1}{2\omega}\sqrt{\omega^{4}+m^{4}_{\emph{K}}+m^{4}_{\pi}-2(\omega^{2}m^{2}_{\emph{K}}+\omega^{2}m^{2}_{\pi}+m^{2}_{\emph{K}}m^{2}_{\pi})},\\
&\mid {\overrightarrow{p_{3}}} \mid=\frac{1}{2\omega}\sqrt{M^{4}_{B^0_{s}}+M^{4}_{\psi}+\omega^{4}-2(M^{2}_{B^0_{s}}M^{2}_{\psi}+M^{2}_{B^0_{s}}\omega^{2}+M^{2}_{\psi}\omega^{2})}.
\end{split}
\end{equation}

The terms $\cal{A}_{\emph{0}}$, $\cal{A}_{\|}$, and $\cal{A}_{\bot}$ represent the longitudinal, parallel, and perpendicular polarization amplitudes, respectively. The related expressions are
\begin{equation}
\begin{split}
&\cal{A}_{\emph{0}}=\cal{A}_{\emph{L}},\\
&\cal{A}_{\|}=\sqrt{\emph{2}} \cal{A}_{\emph{N}},\\
&\cal{A}_{\bot}=\sqrt{\emph{2}} \cal{A}_{\emph{T}},
\end{split}
\end{equation}
where the subscripts $\emph{L}$, $\emph{N}$, and $\emph{T}$ denote the longitudinal, normal, and transverse polarizations, respectively. The polarization fraction is defined as
\begin{equation}
f_{i}=\frac{\mid {\cal A}_{i} \mid^2}{\mid {\cal A}_{\emph{0}}\mid^2+\mid {\cal A}_{\|}\mid^2+\mid {\cal A}_{\bot}\mid^2},
\label{eq:exp39}
\end{equation}
with the normalization relation $f_{\emph{0}}+f_{\|}+f_{\bot}=1$.

\section{Decay amplitudes}\label{sec:amp}
Based on the pQCD approach, the decay amplitude of $B^{0}_{s} \rightarrow \psi(2S,1D)\emph{K}^-\pi^+$ is
\begin{equation}
{\cal{A}_{\emph{L},\emph{N},\emph{T}}}=\frac{\emph{G}_{\emph{F}}}{\sqrt{2}}\bigg[V^*_{cb}V_{cd}\big(\emph{a}_{1}\emph{F}^{\emph{LL}}_{\emph{L},\emph{N},\emph{T}}+
\emph{C}_{2}\emph{M}^{\emph{LL}}_{\emph{L},\emph{N},\emph{T}}\big)-V^*_{tb}V_{td}\big(\emph{a}_{2}\emph{F}^{\emph{LL}}_{\emph{L},\emph{N},\emph{T}}+\emph{a}_{3}\emph{F}^{\emph{LR}}_{\emph{L},\emph{N},\emph{T}}
+(\emph{C}_{4}+\emph{C}_{10})\emph{M}^{\emph{LL}}_{\emph{L},\emph{N},\emph{T}}+(\emph{C}_{6}+\emph{C}_{8})\emph{M}^{\emph{SP}}_{\emph{L},\emph{N},\emph{T}}\big)\bigg],
\label{eq:amp}
\end{equation}
where $\emph{F}$ and $\emph{M}$ represent the factorization and non-factorization contributions, respectively. The superscripts $\emph{LL}$ and $\emph{LR}$ denote the weak vertices of the operators, and $\emph{SP}$ is the Fierz transformation of $\emph{LR}$. For the S-wave, the amplitude is only a longitudinal polarization. The total decay amplitudes of the P-wave and the D-wave are decomposed into
\begin{equation}
\cal{A}=\cal{A}_{\emph{L}}+\cal{A}_{\emph{N}}\epsilon_{\emph{T}}\cdot\epsilon_{\emph{3T}}+\emph{i}\cal{A}_{\emph{T}}\epsilon_{\alpha\beta\rho\sigma}
\emph{n}^{\alpha}_{+} \emph{n}^{\beta}_{-} \epsilon^{\rho}_{\emph{T}} \epsilon^{\sigma}_{\emph{3T}}.
\end{equation}

The decay amplitudes of the longitudinal polarization are as follows:
\begin{equation}
\begin{split}
{\emph F}^{\emph{LL}}_{\emph{L}}(\emph{S})=&8\pi\emph{C}_{\emph{F}}f_{\psi}M^{4}_{B^0_{s}}\int^{1}_{0}\emph{d}\emph{x}_{B}\emph{d}\emph{z}\int^\infty_{0}\emph{b}_{B}\emph{b}\emph{d}\emph{b}_{B}\emph{d}\emph{b}\phi_{B_{s}}(\emph{x}_{B},b_{B})\\
& \times \{[((1-\eta)(1+(1-2r^2)z)-r^2) \phi^{0}_{S}+ \sqrt{(1-r^2)\eta}[(1-2z-\eta+2\eta z-(1-2z+2\eta z)r^2)( \phi^{s}_{S}+ \phi^{t}_{S})+2r^2 \phi^{t}_{S}]]\\
& \times \alpha_{s}(t_{a})\exp[-S_{B^{0}_{s}}(t_{a})-S_{M}(t_{a})]S_{t}(z)h_{a}(x_{B}, z, b_{B}, b)\\
& +[[\eta^{2}-\eta+(\eta-x_{B})r^2](1-r^2)\phi^{0}_{S}+2\sqrt{(1-r^2)\eta}[1-\eta-r^2(1-x_{B})]\phi^{s}_{S}]\\
& \times \alpha_{s}(t_{b})\exp[-S_{B^{0}_{s}}(t_{b})-S_{M}(t_{b})]S_{t}(|x_{B}-\eta|)h_{b}(x_{\emph{B}}, z, b_{\emph{B}}, b)\},
\end{split}
\end{equation}
\begin{equation}
\begin{split}
{\emph{F}}^{\emph{LR}}_{\emph{L}}(\emph{S})={\emph{F}}^{\emph{LL}}_{\emph{L}}(\emph{S}),
\end{split}
\end{equation}
\begin{equation}
\begin{split}
{\emph{M}}^{\emph{LL}}_{\emph{L}}(\emph{S})=&\frac{-32\pi\emph{C}_{\emph{F}}M^{4}_{B^{0}_{s}}}{\sqrt{2N_{c}}}\int^{1}_{0}\emph{d}\emph{x}_{B}\emph{d}\emph{z}\emph{d}\emph{x}_{3}\int^\infty_{0}\emph{b}_{B}\emph{b}_{3}\emph{d}\emph{b}_{B}\emph{d}\emph{b}_{3}\phi_{B_{s}}(\emph{x}_{B},b_{B})\\
& \times \{[(1-r^2-\eta)[((1-x_{B}-x_{3})(1-r^2)+\eta((1-2r^2)x_{3}\\
&-1+z-zr^2))\psi^{L}(x_{3},b_{3})+(1-\eta)rr_{c}\psi^{t}(x_{3},b_{3})]\phi^{0}_{S}\\
&+\sqrt{(1-r^2)\eta}[((1-r^2)z+2(1-x_{3})r^2-x_{B}r^2)(1-\eta)\phi^{t}_{S}-((1-r^2)z(1-\eta)+x_{B}r^2)\phi^{s}_{S}]\psi^{L}(x_{3},b_{3})]\\
& \times \alpha_{s}(t_{c})\exp[-S_{B^{0}_{s}}(t_{c})-S_{M}(t_{c})-S_{\psi}(t_{c})]h_{c}(x_{B}, z, x_{3}, b_{B}, b_{3})\\
& +[(1-r^2-\eta)[(x_{B}-z+zr^2-(1+r^2-\eta)x_{3})\psi^{L}(x_{3},b_{3})+(1-\eta)rr_{c} \psi^{t}(x_{3},b_{3})]\phi^{0}_{S}\\
& -\sqrt{(1-r^2)\eta}[((x_{B}r^2-(2x_{3}r^2+(1-r^2)z)(1-\eta))\psi^{L}(x_{3},b_{3})+4(1-\eta)r r_{c}\psi^{t}(x_{3},b_{3}))\phi^{t}_{S}\\
& -(x_{B}r^2+(1-\eta)z(1-r^2))\psi^{L}(x_{3},b_{3})\phi^{s}_{S}]]\\
& \times \alpha_{s}(t_{d})\exp[-S_{B^{0}_{s}}(t_{d})-S_{M}(t_{d})-S_{\psi}(t_{d})]h_{d}(x_{B}, z, x_{3}, b_{B}, b_{3})\},
\end{split}
\end{equation}

\begin{equation}
\begin{split}
{\emph{M}}^{\emph{SP}}_{\emph{L}}(\emph{S})=&\frac{32\pi\emph{C}_{\emph{F}}M^{4}_{B^{0}_{s}}}{\sqrt{2N_{c}}}\int^{1}_{0}\emph{d}\emph{x}_{B}\emph{d}\emph{z}\emph{d}\emph{x}_{3}\int^\infty_{0}\emph{b}_{B}\emph{b}_{3}\emph{d}\emph{b}_{B}\emph{d}\emph{b}_{3}\phi_{B_{s}}(\emph{x}_{B},b_{B})\\
& \times \{[(1-\eta-r^2)[((1-x_{3})(1+r^2-\eta)-x_{B}+z(1-r^2))\psi^{L}(x_{3},b_{3})-(1-\eta)r r_{c}\psi^{t}(x_{3},b_{3})]\phi^{0}_{S}\\
& +\sqrt{(1-r^2)\eta}[((1-\eta)(r^2-1)z-x_{B}r^2)\phi^{s}_{S}\psi^{L}(x_{3},b_{3})\\
&+[((1-\eta)((r^2-1)z-2(1-x_{3})r^2)+x_{B}r^2)\psi^{L}(x_{3},b_{3})+4(1-\eta)rr_{c}\psi^{t}(x_{3},b_{3})]\phi^{t}_{S}]]\\
& \times \alpha_{s}(t_{c})\exp[-S_{B^{0}_{s}}(t_{c})-S_{M}(t_{c})-S_{\psi}(t_{c})]h_{c}(x_{B}, z, x_{3}, b_{B}, b_{3})\\
& +[(1-r^2-\eta)[((x_{B}-z\eta)(1-r^2)+x_{3}(\eta-1+r^2(1-2\eta)))\psi^{L}(x_{3},b_{3})-(1-\eta)rr_{c}
\psi^{t}(x_{3},b_{3})]\phi^{0}_{S}\\
& +\sqrt{(1-r^2)\eta}[((1-r^2)z(1-\eta)+x_{B}r^2)\phi^{s}_{S}+((z+2x_{3}r^{2}-zr^2)(\eta-1)+x_{B}r^2)\phi^{t}_{S}]\psi^{L}(x_{3},b_{3})]\\
& \times \alpha_{s}(t_{d})\exp[-S_{B^{0}_{s}}(t_{d})-S_{M}(t_{d})-S_{\psi}(t_{d})]h_{d}(x_{B}, z, x_{3}, b_{B}, b_{3})\}.
\end{split}
\end{equation}

$\cal{A}_{\emph{L}}(\emph{P})$ and $\cal{A}_{\emph{L}}(\emph{D})$ can be expressed by the following replacement:
\begin{equation}
\begin{split}
&\cal{A}_{\emph{L}}(\emph{P})=\cal{A}_{\emph{L}}(\emph{S})|_{\phi^{\emph{0},\emph{s}}_{\emph{S}}\rightarrow\phi^{\emph{0},\emph{s}}_{\emph{P}},\phi^{\emph{t}}_{\emph{S}}\rightarrow (\emph{1}-\emph{r}^{\emph{2}})\phi^{\emph{t}}_{\emph{P}}},\\
&\cal{A}_{\emph{L}}(\emph{D})=\sqrt{\frac{\emph{2}}{\emph{3}}} \cal{A}_{\emph{L}}(\emph{S})|_{\phi^{\emph{0},\emph{s}}_{\emph{S}}\rightarrow\phi^{\emph{0},\emph{s}}_{\emph{D}},\phi^{\emph{t}}_{\emph{S}}\rightarrow (\emph{1}-\emph{r}^{\emph{2}})\phi^{\emph{t}}_{\emph{D}}}.
\end{split}
\end{equation}

The decay amplitudes of normal polarization are as follows:
\begin{equation}
\begin{split}
{\emph F}^{\emph{LL}}_{\emph{N}}(\emph{P}) =&8\pi\emph{C}_{\emph{F}}f_{\psi}M^{4}_{B^0_{s}}r\int^{1}_{0}\emph{d}\emph{x}_{B}\emph{d}\emph{z}\int^\infty_{0}\emph{b}_{B}\emph{b}\emph{d}\emph{b}_{B}\emph{d}\emph{b}\phi_{B_{s}}(\emph{x}_{B},b_{B})\\
& \times \{[(r^2-1-(1-2z+2zr^2)\eta)\phi^{T}_{P}+\sqrt{(1-r^2)\eta}((zr^2-2-z)\phi^{a}_{P}+z(1-r^2)\phi^{v}_{P})]\\
&\times \alpha_{s}(t_{a})\exp[-S_{B^{0}_{s}}(t_{a})-S_{M}(t_{a})]S_{t}(z)h_{a}(x_{B}, z, b_{B}, b)\\
& -\sqrt{(1-r^2)\eta}[(1+\eta-x_{B}-r^2)\phi^{a}_{P}+(1+x_{B}-\eta-r^2)\phi^{v}_{P}]\\
& \times \alpha_{s}(t_{b})\exp[-S_{B^{0}_{s}}(t_{b})-S_{M}(t_{b})]S_{t}(|x_{B}-\eta|)h_{b}(x_{\emph{B}}, z, b_{\emph{B}}, b)\},
\end{split}
\end{equation}
\begin{equation}
\begin{split}
{\emph{F}}^{\emph{LR}}_{\emph{N}}(\emph{P})={\emph{F}}^{\emph{LL}}_{\emph{N}}(\emph{P}),
\end{split}
\end{equation}
\begin{equation}
\begin{split}
{\emph{M}}^{\emph{LL}}_{\emph{N}}(\emph{P})=&\frac{-64\pi\emph{C}_{\emph{F}}M^{4}_{B^{0}_{s}}}{\sqrt{2N_{c}}}\int^{1}_{0}\emph{d}\emph{x}_{B}\emph{d}\emph{z}\emph{d}\emph{x}_{3}\int^\infty_{0}\emph{b}_{B}\emph{b}_{3}\emph{d}\emph{b}_{B}\emph{d}\emph{b}_{3}\phi_{B_{s}}(\emph{x}_{B},b_{B})\\
& \times \{[(x_{3}-x_{B}+z\eta-x_{3}\eta)r\psi^{V}(x_{3},b_{3})-(1-\eta)r_{c}\psi^{T}(x_{3},b_{3})]\phi^{T}_{P}\\
& +\sqrt{(1-r^2)\eta}[(x_{B}-x_{3}-z+x_{3}\eta)r\psi^{V}(x_{3},b_{3})+(1-\eta)r_{c}\psi^{T}(x_{3},b_{3})]\phi^{a}_{P}\}\\
& \times \alpha_{s}(t_{d})\exp[-S_{B^{0}_{s}}(t_{d})-S_{M}(t_{d})-S_{\psi}(t_{d})]h_{d}(x_{B}, z, x_{3}, b_{B}, b_{3}),
\end{split}
\end{equation}
\begin{equation}
\begin{split}
{\emph{M}}^{\emph{SP}}_{\emph{N}}(\emph{P})=-{\emph{M}}^{\emph{LL}}_{\emph{N}}(\emph{P}).
\end{split}
\end{equation}

$\cal{A}_{\emph{N}}(\emph{D})$ can be expressed by the following replacement:
\begin{equation}
\begin{split}
\cal{A}_{\emph{N}}(\emph{D})=\sqrt{\frac{\emph{1}}{\emph{2}}} \cal{A}_{\emph{N}}(\emph{P})|_{\phi^{\emph{T},\emph{a},\emph{v}}_{\emph{P}}\rightarrow\phi^{\emph{T},\emph{a},\emph{v}}_{\emph{D}}}.
\end{split}
\end{equation}

The decay amplitudes of transverse polarization are as follows:
\begin{equation}
\begin{split}
{\emph F}^{\emph{LL}}_{\emph{T}}(\emph{P}) =&8\pi\emph{C}_{\emph{F}}f_{\psi}M^{4}_{B^0_{s}}r\int^{1}_{0}\emph{d}\emph{x}_{B}\emph{d}\emph{z}\int^\infty_{0}\emph{b}_{B}\emph{b}\emph{d}\emph{b}_{B}\emph{d}\emph{b}\phi_{B_{s}}(\emph{x}_{B},b_{B})\\
& \times \{[(r^2-1+(1-2z+2zr^2)\eta)\phi^{T}_{P}+\sqrt{(1-r^2)\eta}((zr^2-2-z)\phi^{v}_{P}+z(1-r^2)\phi^{a}_{P})]\\
&\times \alpha_{s}(t_{a})\exp[-S_{B^{0}_{s}}(t_{a})-S_{M}(t_{a})]S_{t}(z)h_{a}(x_{B}, z, b_{B}, b)\\
& -\sqrt{(1-r^2)\eta}[(1+\eta-x_{B}-r^2)\phi^{v}_{P}+(1+x_{B}-\eta-r^2)\phi^{a}_{P}]\\
& \times \alpha_{s}(t_{b})\exp[-S_{B^{0}_{s}}(t_{b})-S_{M}(t_{b})]S_{t}(|x_{B}-\eta|)h_{b}(x_{\emph{B}}, z, b_{\emph{B}}, b)\},
\end{split}
\end{equation}
\begin{equation}
\begin{split}
{\emph{F}}^{\emph{LR}}_{\emph{T}}(\emph{P})={\emph{F}}^{\emph{LL}}_{\emph{T}}(\emph{P}),
\end{split}
\end{equation}
\begin{equation}
\begin{split}
{\emph{M}}^{\emph{LL}}_{\emph{T}}(\emph{P})=&\frac{-64\pi\emph{C}_{\emph{F}}M^{4}_{B^{0}_{s}}}{\sqrt{2N_{c}}}\int^{1}_{0}\emph{d}\emph{x}_{B}\emph{d}\emph{z}\emph{d}\emph{x}_{3}\int^\infty_{0}\emph{b}_{B}\emph{b}_{3}\emph{d}\emph{b}_{B}\emph{d}\emph{b}_{3}\phi_{B_{s}}(\emph{x}_{B},b_{B})\\
& \times \{[(x_{3}-x_{B}-z\eta-x_{3}\eta)r\psi^{V}(x_{3},b_{3})-(1-\eta)r_{c}\psi^{T}(x_{3},b_{3})]\phi^{T}_{P}\\
& +\sqrt{(1-r^2)\eta}[(x_{B}-x_{3}-z+x_{3}\eta)r\psi^{V}(x_{3},b_{3})+(1-\eta)r_{c}\psi^{T}(x_{3},b_{3})]\phi^{v}_{P}\}\\
& \times \alpha_{s}(t_{d})\exp[-S_{B^{0}_{s}}(t_{d})-S_{M}(t_{d})-S_{\psi}(t_{d})]h_{d}(x_{B}, z, x_{3}, b_{B}, b_{3}),
\end{split}
\end{equation}
\begin{equation}
\begin{split}
{\emph{M}}^{\emph{SP}}_{\emph{T}}(\emph{P})=-{\emph{M}}^{\emph{LL}}_{\emph{T}}(\emph{P}).
\end{split}
\end{equation}

$\cal{A}_{\emph{T}}(\emph{D})$ can be expressed by the following replacement:
\begin{equation}
\begin{split}
\cal{A}_{\emph{T}}(\emph{D})=\sqrt{\frac{\emph{1}}{\emph{2}}} \cal{A}_{\emph{T}}(\emph{P})|_{\phi^{\emph{T},\emph{a},\emph{v}}_{\emph{P}}\rightarrow\phi^{\emph{T},\emph{a},\emph{v}}_{\emph{D}}}.
\end{split}
\end{equation}

The mass ratio $r_{c}=\frac{m_{c}}{M_{B^{0}_{s}}}$ and the group factor $\emph{C}_{\emph{F}}=\frac{4}{3}$. The expressions for the Sudakov exponents $S_{B^{0}_{s}}(t)$, $S_{M}(t)$, and $S_{\psi}(t)$, the threshold resummation factor $S_{t}(x)$, the scattering kernel functions $h_{i}(i=a, b, c, d)$, and the hard scales $t_{i}$ have been given in the APPENDIX.

Vertex correction has been performed on the factorization diagrams in this work. According to the NDR scheme~\cite{8319141999,5913132000,6753332003}, the relevant Wilson coefficients are expressed as
\begin{equation}
\begin{split}
&{\emph{a}}_{1}(\emph{S})=\emph{C}_{1}+\frac{\emph{C}_{2}}{N_{c}}+\frac{\alpha_{s}}{9\pi}\emph{C}_{2}[-18-12\textmd{ln}\bigg(\frac{\mu}{\emph{m}_{b}}\bigg)+\emph{f}_{I}+(1-r^{2})\emph{g}_{I}],\\
&{\emph{a}}_{2}(\emph{S})=\emph{C}_{3}+\frac{\emph{C}_{4}}{N_{c}}+\emph{C}_{9}+\frac{\emph{C}_{10}}{N_{c}}+\frac{\alpha_{s}}{9\pi}(\emph{C}_{4}+\emph{C}_{10})[-18-12\textmd{ln}\bigg(\frac{\mu}{\emph{m}_{b}}\bigg)+\emph{f}_{I}+(1-r^{2})\emph{g}_{I}],\\
&{\emph{a}}_{3}(\emph{S})=\emph{C}_{5}+\frac{\emph{C}_{6}}{N_{c}}+\emph{C}_{7}+\frac{\emph{C}_{8}}{N_{c}}+\frac{\alpha_{s}}{9\pi}(\emph{C}_{6}+\emph{C}_{8})[6+12\textmd{ln}\bigg(\frac{\mu}{\emph{m}_{b}}\bigg)-\emph{f}_{I}-(1-r^{2})\emph{g}_{I}],\\
\end{split}
\end{equation}

\begin{equation}
\begin{split}
&{\emph{a}}_{1}(\emph{P,D})=\emph{C}_{1}+\frac{\emph{C}_{2}}{N_{c}}+\frac{\alpha_{s}}{9\pi}\emph{C}_{2}[-18-12\textmd{ln}\bigg(\frac{\mu}{\emph{m}_{b}}\bigg)+f^{\emph{h}}],\\
&{\emph{a}}_{2}(\emph{P,D})=\emph{C}_{3}+\frac{\emph{C}_{4}}{N_{c}}+\emph{C}_{9}+\frac{\emph{C}_{10}}{N_{c}}+\frac{\alpha_{s}}{9\pi}(\emph{C}_{4}+\emph{C}_{10})[-18-12\textmd{ln}\bigg(\frac{\mu}{\emph{m}_{b}}\bigg)+f^{\emph{h}}],\\
&{\emph{a}}_{3}(\emph{P,D})=\emph{C}_{5}+\frac{\emph{C}_{6}}{N_{c}}+\emph{C}_{7}+\frac{\emph{C}_{8}}{N_{c}}+\frac{\alpha_{s}}{9\pi}(\emph{C}_{6}+\emph{C}_{8})[6+12\textmd{ln}\bigg(\frac{\mu}{\emph{m}_{b}}\bigg)-f^{\emph{h}}].\\
\end{split}
\end{equation}
The renormalization scale, $\mu$, has been selected to be of the order of $\emph{m}_{b}$. The Wilson coefficients ${\emph{a}}_{1,2,3}(\emph{S})$ were applied to the decay amplitude $\cal{A}(\emph{S})$ with only longitudinal polarization, and the hard scattering functions, $\emph{f}_{I}$ and $\emph{g}_{I}$, are given in Ref.~\cite{630740112001}. Meanwhile, the Wilson coefficients ${\emph{a}}_{1,2,3}(\emph{P,D})$ were applied to the decay amplitudes $\cal{A}(\emph{P,D})$ with both longitudinal and transverse polarizations, the hard scattering function, $f^{\emph{h}}$, comes from the vertex corrections, and the superscript $\emph{h}$ denotes the polarization state: $\emph{h}=0$ for the helicity $0$ state, whereas $\emph{h}=\pm$ for the helicity $\pm$ states. The expressions for $f^{0}$ and $f^{\pm}$ can be found in Ref.~\cite{650940232002}.

According to the 2S-1D mixing scheme, the decay amplitudes of $B^{0}_{s} \rightarrow \psi(3686,3770)\emph{K}^{-}\pi^+$ can be constructed as
\begin{equation}
{\cal A}(B^{0}_{s}\rightarrow \psi(3686)\emph{K}^{-}\pi^{+})=\sin\theta {\cal A}(B^{0}_{s}\rightarrow \psi(1D)\emph{K}^{-}\pi^+) + \cos\theta {\cal A}(B^{0}_{s}\rightarrow \psi(2S)\emph{K}^{-}\pi^+),
\label{eq:exp40}
\end{equation}
\begin{equation}
{\cal A}(B^{0}_{s}\rightarrow \psi(3770)\emph{K}^{-}\pi^{+})=\cos\theta {\cal A}(B^{0}_{s}\rightarrow \psi(1D)\emph{K}^{-}\pi^+) - \sin\theta {\cal A}(B^{0}_{s}\rightarrow \psi(2S)\emph{K}^{-}\pi^+).
\label{eq:exp41}
\end{equation}


\section{Numerical results and discussions}\label{sec:numer}

\begin{table}[htbp]
\centering
\caption{Various parameters used in the calculation~\cite{770540032008,8083c012020,817442021}.}
\label{tab1}
\begin{tabular*}{\columnwidth}{@{\extracolsep{\fill}}lllll@{}}
\hline
\hline
\\
Masses   &$M_{B^{0}_{s}}=5.367$ {\rm GeV}     &$M_{\psi_{(2S)}}=3.686$ {\rm GeV}       &$M_{\psi_{(1D)}}=3.77$ {\rm GeV}  \vspace{1ex} \\
                                  &$m_{b}=4.75$ {\rm GeV}                  &$m_{c}=1.4$ {\rm GeV}       &$m_{\emph{K}}=0.494$ {\rm GeV}   \vspace{1ex} \\
                                  &$m_{\pi}=0.140$ {\rm GeV}  \vspace{1ex} \\
Decay constants   &$f_{B^0_{s}}=227.2 \pm 3.4$ {\rm MeV}   &$f_{\psi_{(2S)}}=296^{+3}_{-2}$ {\rm MeV}                    &$ f_{\psi_{(1D)}}=45.8$ {\rm MeV} \vspace{1ex} \\
Lifetime of meson               &$\tau_{B^{0}_{s}}=1.509$ {\rm ps}     \vspace{1ex} \\
Wolfenstein parameters           &$\emph{A}=0.836\pm0.015$              &$\lambda=0.22453\pm0.00044$   &$\bar{\eta}=0.355^{+0.012}_{-0.011}$ \vspace{1ex} \\
                                   &$\bar{\rho}=0.122^{+0.018}_{-0.017}$  \vspace{1ex} \\
\hline
\end{tabular*}
\end{table}

\begin{table}[htbp]
\centering
\caption{Pole masses and widths for the different resonances~\cite{7810192018}.}
\label{tab2}
\begin{tabular*}{\columnwidth}{@{\extracolsep{\fill}}c c c@{}}
\hline
\hline
$\textmd{Resonance}$  & $\textmd{Mass}$  & $\textmd{Width}$  \\
\hline
  \\
$\emph{K}^{*}(892)^{0}$  & $895.55\pm0.20$ {\rm MeV}  & $47.3\pm0.5$ {\rm MeV} \vspace{1ex}  \\
$\emph{K}^{*}(1410)^{0}$  & $1414\pm15$ {\rm MeV}  & $232\pm21$ {\rm MeV} \vspace{1ex}   \\
$\emph{K}^{*}_{0}(1430)^{0}$  & $1425\pm50$ {\rm MeV}  & $270\pm80$ {\rm MeV} \vspace{1ex}    \\
$\emph{K}^{*}_{2}(1430)^{0}$  & $1432.4\pm1.3${\rm MeV}  & $109\pm5$ {\rm MeV} \vspace{1ex}    \\
$\emph{K}^{*}(1680)^{0}$  & $1717\pm27$ {\rm MeV}  & $322\pm110$ {\rm MeV} \vspace{1ex}   \\
\hline
\hline
\end{tabular*}
\end{table}

The parameters used in the calculation have been presented in Table \ref{tab1}, which include the masses of the involved mesons, their decay constants, the lifetime of the $B^{0}_{s}$ meson, and the Wolfenstein parameters. The pole masses of the quarks were adopted in this study~\cite{817442021}.

The data in Table \ref{tab2} have been taken from Ref.~\cite{7810192018}, the relevant information that should be considered in the study for the S, P, and D-wave resonances are contained in the table. In this work, the dynamic limit of the invariant mass of the resonance is $m_{\emph{K}}+m_{\pi}< \omega < {M_{B_{s}^{0}}-M_{\psi}}$. In addition, although the mass of the $\overline{\emph{K}}^{*}(1680)^{0}$ resonance exceeds the upper limit, its decay channels should be considered in the study because of its large width($\Gamma_{\emph{K}^{*}(1680)^{0}}=322\pm110$ {\rm MeV}).

The decay branching ratios of the $\overline{\emph{K}}^{*}_{0}(1430)^{0}$ resonance of the S-wave were first calculated and the results obtained have been given in Table \ref{tab3}. The errors were derived from the shape parameter, $\omega_{B_{s}}$, in the wave function of the $B^{0}_{s}$ meson, the Gegenbauer moments in the DAs of the kaon-pion pair, and the hard scale $\emph{t}(0.9\emph{t}\thicksim1.1\emph{t})$, respectively. The errors in the following tables were analyzed in the same order.

Next, the resonances of the P-wave were calculated considering $\overline{\emph{K}}^{*}(892)^{0}$, $\overline{\emph{K}}^{*}(1410)^{0}$, and $\overline{\emph{K}}^{*}(1680)^{0}$, and the results thus obtained have been given in Table \ref{tab4}. The experimental measurement data $\cal{B}$$\big(B^{0}_{s} \rightarrow \psi(2S)\overline{\emph{K}}^{*}(892)^{0}(\rightarrow\emph{K}^-\pi^+)\big)=(2.2\pm0.3)\times10^{-5}$ was taken from the article of Zyla et al.~\cite{8083c012020}. Our pQCD prediction agrees well with it within errors.

Finally, the contributions of the $\overline{\emph{K}}^{*}_{2}(1430)^{0}$ intermediate resonance of the D-wave were considered and the calculation results have been presented in Table \ref{tab5}.

\begin{table}[htbp]
\centering
\caption{
Branching ratios of the S-wave resonance in the quasi-two-body decays $B^{0}_{s} \rightarrow
\psi(2S,1D)\overline{\emph{K}}^{*0}(\rightarrow\emph{K}^{-}\pi^{+})$ calculated using the pQCD factorization approach.}
\label{tab3}
\begin{tabular*}{\columnwidth}{@{\extracolsep{\fill}}ccc@{}}
\hline
\hline
$\textmd{Decay mode}$  & $\textmd{pQCD prediction}$  & $\textmd{Experimental data}$   \\
\hline
  \\
$B^{0}_{s} \rightarrow \psi(2S)\overline{\emph{K}}^{*}_{0}(1430)^{0}(\rightarrow\emph{K}^-\pi^+)$  & $3.94^{+1.82+0.56+0.11}_{-1.16-0.49-0.07}\times10^{-6}$  & $\cdots$ \vspace{1ex}  \\
$B^{0}_{s} \rightarrow \psi(1D)\overline{\emph{K}}^{*}_{0}(1430)^{0}(\rightarrow\emph{K}^-\pi^+)$  & $1.43^{+0.53+0.07+0.03}_{-0.37-0.07-0.02}\times10^{-6}$  & $\cdots$ \vspace{1ex}  \\
\hline
\hline
\end{tabular*}
\end{table}

\begin{table}[htbp]
\centering
\caption{Branching ratios of the P-wave resonances in the quasi-two-body decays $B^{0}_{s} \rightarrow
\psi(2S,1D)\overline{\emph{K}}^{*0}(\rightarrow\emph{K}^{-}\pi^{+})$ calculated using the pQCD factorization approach.}
\label{tab4}
\begin{tabular*}{\columnwidth}{@{\extracolsep{\fill}}cccc@{}}
\hline
\hline
$\textmd{Decay mode}$   & $ $   & $\textmd{pQCD prediction}$  & $\textmd{Experimental data}$   \\
\hline
\\
$B^{0}_{s} \rightarrow \psi(2S)\overline{\emph{K}}^{*}(892)^{0}(\rightarrow\emph{K}^-\pi^+)$  & $\cal{B}$ $(10^{-5})$ & $2.71^{+1.16+1.03+0.11}_{-0.83-0.84-0.07}$  & $2.20\pm0.33$ \vspace{1ex}  \\
& $f_{0}$ $(\%)$    &  $43.2^{+23.2+9.6+1.5}_{-16.2-8.5-0.7}$  & $52.0\pm6.0$ \vspace{1ex} \\
& $f_{\|}$ $(\%)$    &  $27.7^{+8.5+15.1+1.1}_{-6.6-11.8-0.7}$  & $\cdots$ \vspace{1ex} \\
& $f_{\bot}$ $(\%)$    &  $29.1^{+11.1+13.3+1.5}_{-7.7-10.7-1.1}$  & $\cdots$ \vspace{1ex} \\
\hline
\\$B^{0}_{s} \rightarrow \psi(2S)\overline{\emph{K}}^{*}(1410)^{0}(\rightarrow\emph{K}^-\pi^+)$  & $\cal{B}$ $(10^{-7})$ & $5.19^{+1.69+2.11+0.21}_{-1.36-1.18-0.11}$  & $\cdots$ \vspace{1ex}  \\
& $f_{0}$ $(\%)$    &  $44.1^{+16.6+9.1+1.3}_{-12.9-7.9-0.6}$  & $\cdots$ \vspace{1ex} \\
& $f_{\|}$ $(\%)$    &  $26.8^{+7.3+15.2+1.0}_{-6.2-10.8-0.6}$  & $\cdots$ \vspace{1ex} \\
& $f_{\bot}$ $(\%)$    &  $29.1^{+8.7+16.4+1.7}_{-7.1-4.0-1.0}$  & $\cdots$ \vspace{1ex} \\
\hline
\\$B^{0}_{s} \rightarrow \psi(2S)\overline{\emph{K}}^{*}(1680)^{0}(\rightarrow\emph{K}^-\pi^+)$  & $\cal{B}$ $(10^{-7})$ & $2.33^{+0.80+0.92+0.10}_{-0.58-0.56-0.05}$  & $\cdots$ \vspace{1ex}  \\
& $f_{0}$ $(\%)$    &  $44.2^{+17.2+9.4+1.3}_{-12.0-8.2-0.9}$  & $\cdots$ \vspace{1ex} \\
& $f_{\|}$ $(\%)$    &  $26.6^{+8.2+15.9+0.9}_{-6.0-11.2-0.4}$  & $\cdots$ \vspace{1ex} \\
& $f_{\bot}$ $(\%)$    &  $29.2^{+9.0+14.2+2.1}_{-7.9-4.7-0.9}$  & $\cdots$ \vspace{1ex} \\
\hline
\\$B^{0}_{s} \rightarrow \psi(2S)(\emph{K}^-\pi^+)_{\emph{P}}$  & $\cal{B}$ $(10^{-5})$ & $2.96^{+1.26+1.17+0.13}_{-0.91-0.95-0.09}$  & $\cdots$ \vspace{1ex}  \\
& $f_{0}$ $(\%)$    &  $42.9^{+22.6+11.5+1.7}_{-15.5-9.8-0.7}$  & $\cdots$ \vspace{1ex} \\
& $f_{\|}$ $(\%)$    &  $27.6^{+9.5+15.2+1.4}_{-7.4-11.8-1.0}$  & $\cdots$ \vspace{1ex} \\
& $f_{\bot}$ $(\%)$    &  $29.5^{+10.5+12.8+1.4}_{-7.8-10.5-1.4}$  & $\cdots$ \vspace{1ex} \\
\hline
\\$B^{0}_{s} \rightarrow \psi(1D)\overline{\emph{K}}^{*}(892)^{0}(\rightarrow\emph{K}^-\pi^+)$  & $\cal{B}$ $(10^{-6})$ & $5.41^{+1.57+4.55+0.12}_{-1.16-2.62-0.05}$  & $\cdots$ \vspace{1ex}  \\
& $f_{0}$ $(\%)$    &  $9.6^{+3.1+1.1+0.4}_{-2.8-0.9-0.2}$  & $\cdots$ \vspace{1ex} \\
& $f_{\|}$ $(\%)$    &  $46.8^{+13.1+40.9+0.9}_{-10.0-23.7-0.4}$  & $\cdots$ \vspace{1ex} \\
& $f_{\bot}$ $(\%)$    &  $43.6^{+12.8+42.1+0.9}_{-8.8-24.0-0.4}$  & $\cdots$ \vspace{1ex} \\
\hline
\\$B^{0}_{s} \rightarrow \psi(1D)\overline{\emph{K}}^{*}(1410)^{0}(\rightarrow\emph{K}^-\pi^+)$  & $\cal{B}$ $(10^{-8})$ & $7.46^{+2.38+7.02+0.17}_{-1.80-3.98-0.12}$  & $\cdots$ \vspace{1ex}  \\
& $f_{0}$ $(\%)$    &  $10.5^{+5.5+3.6+0.4}_{-4.0-2.0-0.3}$  & $\cdots$ \vspace{1ex} \\
& $f_{\|}$ $(\%)$    &  $47.5^{+13.7+44.9+1.1}_{-10.5-24.9-0.8}$  & $\cdots$ \vspace{1ex} \\
& $f_{\bot}$ $(\%)$    &  $42.0^{+12.7+45.6+0.8}_{-9.7-26.4-0.5}$  & $\cdots$ \vspace{1ex} \\
\hline
\\$B^{0}_{s} \rightarrow \psi(1D)\overline{\emph{K}}^{*}(1680)^{0}(\rightarrow\emph{K}^-\pi^+)$  & $\cal{B}$ $(10^{-8})$ & $2.72^{+0.89+2.54+0.07}_{-0.65-1.44-0.06}$  & $\cdots$ \vspace{1ex}  \\
& $f_{0}$ $(\%)$    &  $10.3^{+5.9+3.3+0.4}_{-4.0-2.2-0.4}$  & $\cdots$ \vspace{1ex} \\
& $f_{\|}$ $(\%)$    &  $47.4^{+14.0+44.9+1.1}_{-10.3-24.6-1.1}$  & $\cdots$ \vspace{1ex} \\
& $f_{\bot}$ $(\%)$    &  $42.3^{+12.9+45.2+1.1}_{-9.6-26.1-0.7}$  & $\cdots$ \vspace{1ex} \\
\hline
\\$B^{0}_{s} \rightarrow \psi(1D)(\emph{K}^-\pi^+)_{\emph{P}}$  & $\cal{B}$ $(10^{-6})$ & $5.79^{+1.69+4.62+0.11}_{-1.27-2.72-0.05}$  & $\cdots$ \vspace{1ex}  \\
& $f_{0}$ $(\%)$    &  $10.0^{+3.8+1.0+0.3}_{-2.6-0.7-0.2}$  & $\cdots$ \vspace{1ex} \\
& $f_{\|}$ $(\%)$    &  $46.6^{+13.1+39.0+0.9}_{-10.2-22.8-0.3}$  & $\cdots$ \vspace{1ex} \\
& $f_{\bot}$ $(\%)$    &  $43.4^{+12.3+39.7+0.7}_{-9.2-23.5-0.3}$  & $\cdots$ \vspace{1ex} \\
\hline
\hline
\end{tabular*}
\end{table}

\begin{table}[htbp]
\centering
\caption{Branching ratios of the D-wave resonance in the quasi-two-body decays $B^{0}_{s} \rightarrow \psi(2S,1D)\overline{\emph{K}}^{*0}(\rightarrow\emph{K}^{-}\pi^{+})$ calculated using the pQCD factorization approach.}
\label{tab5}
\begin{tabular*}{\columnwidth}{@{\extracolsep{\fill}}cccc@{}}
\hline
\hline
$\textmd{Decay mode}$   & $ $   & $\textmd{pQCD prediction}$  & $\textmd{Experimental data}$   \\
\hline
\\$B^{0}_{s} \rightarrow \psi(2S)\overline{\emph{K}}^{*}_{2}(1430)^{0}(\rightarrow\emph{K}^-\pi^+)$  & $\cal{B}$ $(10^{-6})$ & $3.17^{+1.18+1.90+0.11}_{-0.92-1.46-0.06}$  & $\cdots$ \vspace{1ex}  \\
& $f_{0}$ $(\%)$    &  $39.4^{+15.1+24.6+0.6}_{-11.7-18.3-0.3}$  & $\cdots$ \vspace{1ex} \\
& $f_{\|}$ $(\%)$    &  $33.1^{+12.0+19.2+1.6}_{-9.5-15.1-0.9}$  & $\cdots$ \vspace{1ex} \\
& $f_{\bot}$ $(\%)$    &  $27.5^{+10.1+16.1+1.3}_{-7.9-12.6-0.6}$  & $\cdots$ \vspace{1ex} \\
\hline
\\$B^{0}_{s} \rightarrow \psi(1D)\overline{\emph{K}}^{*}_{2}(1430)^{0}(\rightarrow\emph{K}^-\pi^+)$  & $\cal{B}$ $(10^{-7})$ & $1.46^{+0.41+0.62+0.04}_{-0.31-0.48-0.03}$  & $\cdots$ \vspace{1ex}  \\
& $f_{0}$ $(\%)$    &  $13.0^{+8.9+11.0+0.7}_{-5.5-8.2-0.7}$  & $\cdots$ \vspace{1ex} \\
& $f_{\|}$ $(\%)$    &  $30.1^{+6.2+9.6+0.7}_{-4.8-6.2-0.0}$  & $\cdots$ \vspace{1ex} \\
& $f_{\bot}$ $(\%)$    &  $56.9^{+13.0+21.9+1.4}_{-11.0-18.5-1.4}$  & $\cdots$ \vspace{1ex} \\
\hline
\hline
\end{tabular*}
\end{table}

The theoretical prediction for the branching ratio of the $B^{0}_{s} \rightarrow \psi(2S)\emph{K}^-\pi^+$ decay is $3.67^{+1.56+1.42+0.15}_{-1.12-1.15-0.10}\times10^{-5}$ in this work, which includes contributions from the intermediate resonances of the S, P, and D-wave. This result is consistent with the latest experimental data $(3.1\pm0.4)\times10^{-5}$\cite{8083c012020} within errors. From the numerical results, it has been observed that $\overline{\emph{K}}^{*}(892)^{0}$ is the main contributor to the $B^{0}_{s} \rightarrow \psi(2S)(\emph{K}^-\pi^+)_{\emph{P}}$ decay, accounting for approximately $91.55\%$, whereas the contributions of the $\overline{\emph{K}}^{*}(1410)^{0}$ and $\overline{\emph{K}}^{*}(1680)^{0}$ resonances account for $1.75\%$ and $0.79\%$, respectively. Further, the interference contribution of the three resonances amount to roughly $5.91\%$. The $\overline{\emph{K}}^{*}(892)^{0}$ resonance is also the main source for the $B^{0}_{s} \rightarrow \psi(1D)(\emph{K}^-\pi^+)_{\emph{P}}$ decay, accounting for approximately $93.44\%$, whereas the $\overline{\emph{K}}^{*}(1410)^{0}$ and $\overline{\emph{K}}^{*}(1680)^{0}$ resonances account for $1.29\%$ and $0.47\%$, respectively. In addition, the interference contribution amount to approximately $4.80\%$. Referring to Table \ref{tab4}, the branching ratios of $\psi\overline{\emph{K}}^{*}(1410)^{0}$ and $\psi\overline{\emph{K}}^{*}(1680)^{0}$ decay modes are of the same order, attributable to the large width of the $\overline{\emph{K}}^{*}(1680)^{0}$ resonance.

In comparison, it has been found that the branching ratio of the $\psi(2S)$ decay channel is $2.76$ times that of the $\psi(1D)$ decay channel of the S-wave. Furthermore, the branching ratios of the $\psi(2S)$ decay modes of the P-wave and D-wave are $5.01\sim8.57$ and $21.71$ times larger than those of the $\psi(1D)$ decay modes, respectively. In our calculation, the main contributions of the $\psi(2S)$ and $\psi(1D)$ decay modes of the S-wave were the non-factorized diagrams, the amplitudes of which are slightly affected by the wave functions changing from $\psi(2S)$ to $\psi(1D)$, thus leading to only a small gap to appear between the branching ratios of the two decay modes of the S-wave. However, the amplitudes of the P, D-wave decay channels are dominated by the factorized diagrams, especially the D-wave decay channels, which are significantly affected by the change in the decay constant from $f_{\psi(2S)}$ to $f_{\psi(1D)}$. Thus, a large gap can be observed between the branching ratios of the $\psi(2S)$ and $\psi(1D)$ decay modes. As mentioned above, the different effects of the factorized and non-factorized diagrams in the decay modes of the S, P, and D-wave might be related to the differences in the wave function models about the scalar, vector, and tensor mesons.

In our study, the main uncertainty in the S-wave decay modes comes from the shape parameter $\omega_{B_{s}}$. For the $\psi(2S)$ decay modes of the P-wave, the errors from the shape parameter and the Gegenbauer moments are very close, whereas the maximum error term for the D-wave decay modes is from the Gegenbauer moments. These differences can be interpreted as the range of the values of the Gegenbauer moments of the P-wave kaon-pion DAs to be larger than that of the S-wave kaon-pion DAs (for example, $\emph{B}_{1}=-0.57\pm0.13$ and $\emph{B}_{3}=-0.42\pm0.22$ for the S-wave, and $a^{\|}_{1\emph{K}^{*}}=0.2\pm0.2$ and $a^{\|}_{2\emph{K}^{*}}=0.5\pm0.5$ for the P-wave), and the single Gegenbauer moment of the D-wave dominates the twist-2 as well as twist-3 DAs in the corresponding polarization case. The error caused by the hard scale, $t$, is the smallest among the three error terms, attributable to the selected range$(0.9t-1.1t)$.

The polarization fractions are defined by Eq.~(\ref{eq:exp39}), and they have been listed in Tables \ref{tab4} and \ref{tab5}. For the P-wave $\psi(2S)$ decay mode, the longitudinal polarization fraction is approximately $43\%$, whereas in the $\psi(1D)$ mode, it is about $10\%$, with parallel and vertical fractions being approximately equal in both modes. For the D-wave $\psi(2S)$ decay mode, the three polarization fractions are roughly at the same level of approximately $33\%$ but they are distinctly different in the $\psi(1D)$ mode. We expect additional abundant and detailed data to be obtained from future experiments so that our theoretical predictions can be accurately verified and more systematic analysis for $B^{0}_{s} \rightarrow \psi(2S,1D)\overline{\emph{K}}^{*}(\rightarrow\emph{K}^-\pi^+)$ decays can be performed.

From the experimental data, the relative fraction between the branching ratios has been obtained to be~\cite{7474842015}
\begin{equation}
\begin{split}
\frac{{\cal B}(\overline{B}^0_{s}\rightarrow\psi(2S)\emph{K}^{*}(892)^{0})}{{\cal B}({B^0\rightarrow\psi(2S)\emph{K}}^{*}(892)^{0})}=5.58\pm0.57(\textmd{stat})
\pm0.40({\textmd{syst}})\pm0.32(\emph{f}_{\emph{s}}/\emph{f}_{\emph{d}})\%.
\end{split}
\end{equation}

By comparing the branching ratio of the $B^{0}_{s} \rightarrow \psi(2S)\overline{\emph{K}}^{*}(892)^{0}(\rightarrow\emph{K}^-\pi^+)$ decay, calculated using the pQCD factorization approach, with the pQCD prediction for the $B^{0}\rightarrow \psi(2S)\emph{K}^{*}(892)^{0}(\rightarrow\emph{K}^+\pi^-)$ decay~\cite{440731022020}, we obtain the relative fraction of the theoretical calculation as
\begin{equation}
\begin{split}
\frac{{\cal B}(B^{0}_{s} \rightarrow \psi(2S)\overline{\emph{K}}^{*}(892)^{0}(\rightarrow\emph{K}^{-}\pi^{+}))}{{\cal B}(B^{0} \rightarrow \psi(2S)\emph{K}^{*}(892)^{0}(\rightarrow\emph{K}^{+}\pi^{-}))}=8.01\%.
\end{split}
\end{equation}

The discrepancy in the values comes from the vertex correction and the selection of different values for some of the parameters. However, this discrepancy is still within the acceptable limit. The relative fraction results predicted by the theory agree somewhat with the experimental data, which support the pQCD factorization approach and also contribute to the further studies on resonance mesons.

 \begin{figure}[htbp]
 \centering
 \begin{tabular}{l}
 \includegraphics[width=0.8\textwidth]{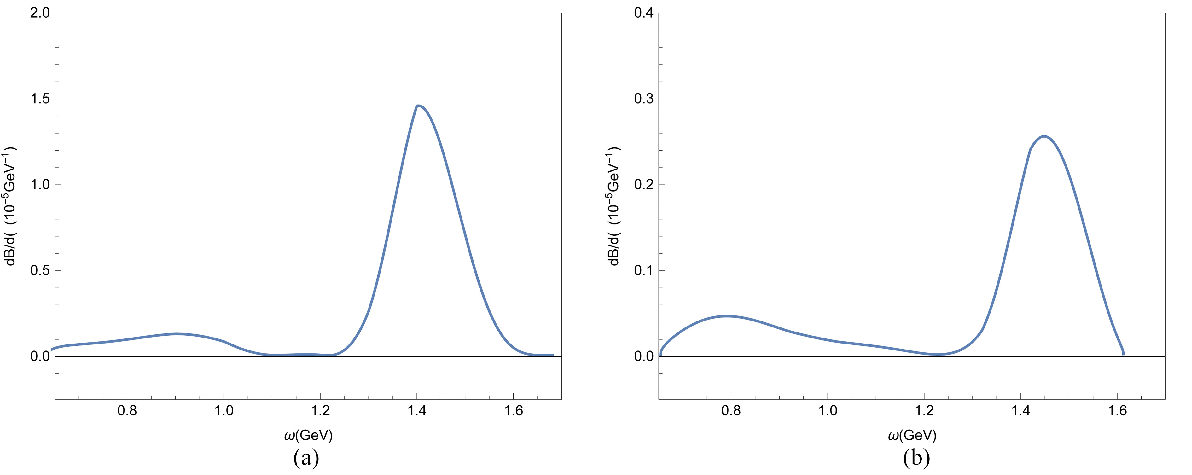}
 \end{tabular}
 \caption {Differential branching ratios of the S-wave for (a) $B^{0}_{s} \rightarrow \psi(2S)\emph{K}^{-}\pi^{+}$ and (b) $B^{0}_{s} \rightarrow \psi(1D)\emph{K}^{-}\pi^{+}$.}
   \label{Swave}
 \end{figure}

 \begin{figure}[htbp]
 \centering
 \begin{tabular}{l}
 \includegraphics[width=0.8\textwidth]{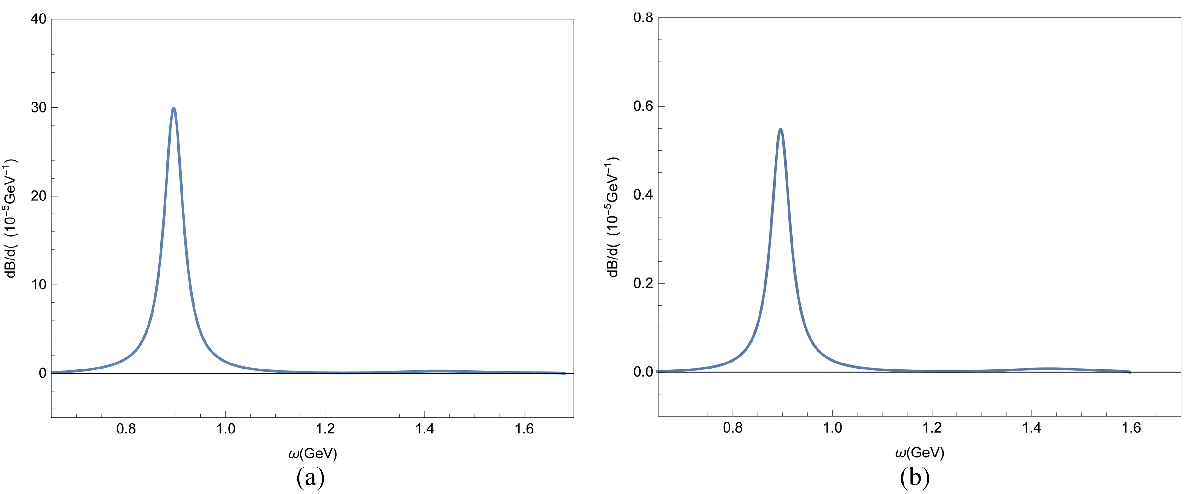}
 \end{tabular}
 \caption {Differential branching ratios of the P-wave for (a) $B^{0}_{s} \rightarrow \psi(2S)\emph{K}^{-}\pi^{+}$ and (b) $B^{0}_{s} \rightarrow \psi(1D)\emph{K}^{-}\pi^{+}$.}
   \label{Pwavetotal}
 \end{figure}

 \begin{figure}[htbp]
 \centering
 \begin{tabular}{l}
 \includegraphics[width=0.8\textwidth]{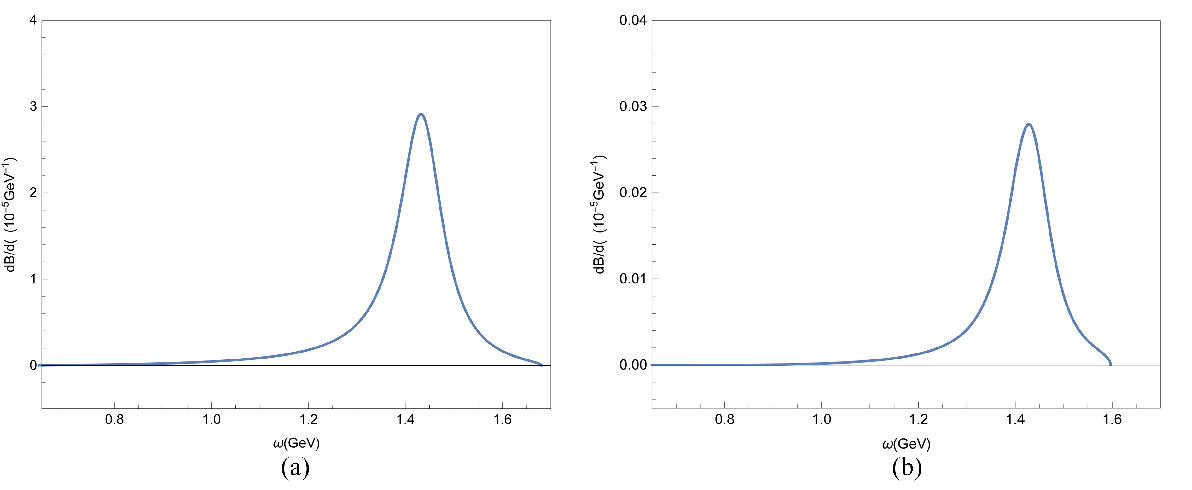}
 \end{tabular}
 \caption {Differential branching ratios of the D-wave for (a) $B^{0}_{s} \rightarrow \psi(2S)\emph{K}^{-}\pi^{+}$ and (b) $B^{0}_{s} \rightarrow \psi(1D)\emph{K}^{-}\pi^{+}$.}
   \label{Dwave}
 \end{figure}

Figs.~\ref{Swave}$-$~\ref{Dwave} depict the function images of the $\omega$ dependence of the differential branching ratios of the S, P, and D-wave of the $B^{0}_{s}\rightarrow\psi(2S,1D)\emph{K}^-\pi^+$ decays, respectively. Fig.~\ref{Swave} shows that a small peak can always be detected near the invariant mass $\omega=0.892${\rm GeV}, which can be attributed to the interference effect of the $\overline{\emph{K}}^{*}(892)^{0}$ resonance on the S-wave. On the other hand, the function images of $\psi(1D)$ mode drop faster at the end than of the $\psi(2S)$ mode due to the difference in the upper limit of their invariant masses $\omega$ of $\emph{K}\pi$. Obviously, the peak values of all function images appear at the pole mass of the corresponding resonance. Therefore, the main part of the branching ratios is in the region around the resonance and almost in the range of $\omega=[m_{\emph{K}^{*}}-\Gamma_{\emph{K}^{*}},m_{\emph{K}^{*}}+\Gamma_{\emph{K}^{*}}]$, the branching ratios of S, P, and D-wave decay modes in this range account for $43.91\%$, $74.73\%$, and $78.68\%$ of the total branching ratios, respectively. The value of $43.91\%$ can be interpreted as the interference effect of the $\overline{\emph{K}}^{*}(892)^{0}$ resonance on the S-wave that is not included.

Using Eqs.~(\ref{eq:exp40}) and ~(\ref{eq:exp41}),the branching ratios of the $B^{0}_{s} \rightarrow \psi(3686,3770)\emph{K}^{-}\pi^{+}$ decays were obtained using the fitting scheme based on the S-D mixing mechanism. The calculation results are presented in Tables \ref{tab6} and \ref{tab7}, respectively.

\begin{table}[htbp]
\centering
\caption{Branching ratios of the quasi-two-body decays $B^{0}_{s} \rightarrow \psi(3686)\overline{\emph{K}}^{*0}(\rightarrow\emph{K}^{-}\pi^{+})$ under the 2S-1D mixing mechanism calculated using the pQCD factorization approach. The first three uncertainties have been derived from the uncertainties in the previous tables, whereas the last one has been derived from the mixing angle.}
\label{tab6}
\begin{tabular*}{\columnwidth}{@{\extracolsep{\fill}}cccc@{}}
\hline
\hline
$\textmd{Decay mode}$  & $ $    & $\theta=(-12\pm2)^{\circ}$  & $\theta=(27\pm2)^{\circ}$   \\
\hline
\\$B^{0}_{s} \rightarrow \psi(3686)\overline{\emph{K}}^{*}_{0}(1430)^{0}(\rightarrow\emph{K}^-\pi^+)$  & $\cal{B}$ $(10^{-6})$  & $2.87^{+1.36+0.50+0.09+0.21}_{-0.85-0.43-0.05-0.21}$  & $5.39^{+2.36+0.49+0.07+0.03}_{-1.52-0.44-0.04-0.03}$  \vspace{1ex}  \\
\hline
\\$B^{0}_{s} \rightarrow \psi(3686)\overline{\emph{K}}^{*}(892)^{0}(\rightarrow\emph{K}^-\pi^+)$  & $\cal{B}$ $(10^{-5})$  & $2.26^{+1.06+0.92+0.11+0.08}_{-0.78-0.77-0.07-0.08}$  & $2.95^{+1.18+1.25+0.12+0.07}_{-0.86-1.03-0.08-0.07}$  \vspace{1ex}  \\
& $f_{0}$ $(\%)$    &  $45.6^{+23.9+12.4+1.8+1.8}_{-16.8-11.1-0.9-1.3}$  & $36.3^{+22.0+7.8+1.7+0.7}_{-14.6-7.5-1.0-1.0}$ \vspace{1ex} \\
& $f_{\|}$ $(\%)$    &  $26.1^{+11.5+15.0+1.3+0.9}_{-8.8-12.8-0.9-0.9}$  & $29.8^{+8.1+17.6+1.0+0.7}_{-7.1-13.9-0.7-0.7}$ \vspace{1ex} \\
& $f_{\bot}$ $(\%)$    &  $28.3^{+11.5+13.3+1.8+0.9}_{-8.8-10.2-1.3-1.3}$  & $33.9^{+9.8+16.9+1.4+1.0}_{-7.5-13.6-1.0-0.7}$ \vspace{1ex} \\
\hline
\\$B^{0}_{s} \rightarrow \psi(3686)\overline{\emph{K}}^{*}(1410)^{0}(\rightarrow\emph{K}^-\pi^+)$  & $\cal{B}$ $(10^{-7})$  & $4.16^{+1.41+1.63+0.20+0.16}_{-1.14-0.90-0.11-0.18}$  & $5.66^{+1.73+2.49+0.21+0.09}_{-1.39-1.52-0.11-0.08}$  \vspace{1ex}  \\
& $f_{0}$ $(\%)$    &  $47.4^{+17.5+10.8+1.4+1.7}_{-13.5-9.4-0.7-1.7}$  & $37.6^{+14.8+6.2+1.2+0.7}_{-11.5-5.4-0.5-0.9}$ \vspace{1ex} \\
& $f_{\|}$ $(\%)$    &  $25.2^{+7.9+14.2+1.2+1.0}_{-6.7-10.3-0.7-1.2}$  & $29.9^{+6.7+17.1+0.9+0.5}_{-5.8-12.9-0.5-0.4}$ \vspace{1ex} \\
& $f_{\bot}$ $(\%)$    &  $27.4^{+8.4+14.2+2.2+1.2}_{-7.2-1.9-1.2-1.4}$  & $32.5^{+9.0+20.7+1.6+0.4}_{-7.2-8.5-0.9-0.2}$ \vspace{1ex} \\
\hline
\\$B^{0}_{s} \rightarrow \psi(3686)\overline{\emph{K}}^{*}(1680)^{0}(\rightarrow\emph{K}^-\pi^+)$  & $\cal{B}$ $(10^{-7})$  & $1.60^{+0.56+0.59+0.10+0.06}_{-0.42-0.37-0.05-0.06}$  & $2.18^{+0.68+0.93+0.11+0.05}_{-0.52-0.60-0.06-0.04}$  \vspace{1ex}  \\
& $f_{0}$ $(\%)$    &  $46.3^{+18.1+10.6+1.9+1.3}_{-13.1-9.4-1.3-1.9}$  & $37.2^{+15.1+6.4+1.4+0.9}_{-11.5-5.5-0.9-0.9}$ \vspace{1ex} \\
& $f_{\|}$ $(\%)$    &  $25.0^{+8.1+14.4+1.3+1.3}_{-6.3-10.0-0.7-0.7}$  & $29.4^{+6.9+18.3+1.4+0.5}_{-5.5-12.4-0.5-0.5}$ \vspace{1ex} \\
& $f_{\bot}$ $(\%)$    &  $28.7^{+8.8+11.9+3.1+1.3}_{-6.9-3.8-1.3-1.3}$  & $33.4^{+9.2+17.9+2.3+0.9}_{-6.9-9.6-1.4-0.5}$ \vspace{1ex} \\
\hline
\\$B^{0}_{s} \rightarrow \psi(3686)(\emph{K}^-\pi^+)_{\emph{P}}$  & $\cal{B}$ $(10^{-5})$  & $2.53^{+1.10+0.95+0.11+0.08}_{-0.77-0.75-0.07-0.09}$  & $3.24^{+1.20+1.25+0.12+0.07}_{-0.89-1.03-0.09-0.07}$  \vspace{1ex}  \\
& $f_{0}$ $(\%)$    &  $44.7^{+22.5+11.9+1.6+1.6}_{-14.6-9.9-0.8-1.6}$  & $36.4^{+20.4+7.1+1.5+0.6}_{-13.6-6.8-0.9-0.9}$ \vspace{1ex} \\
& $f_{\|}$ $(\%)$    &  $26.5^{+10.3+13.8+1.2+0.8}_{-7.9-11.1-0.8-0.8}$  & $29.9^{+7.4+16.4+0.9+0.6}_{-6.5-12.7-0.9-0.6}$ \vspace{1ex} \\
& $f_{\bot}$ $(\%)$    &  $28.8^{+10.7+11.9+1.6+0.8}_{-7.9-8.6-1.2-1.2}$  & $33.7^{+9.3+15.7+1.2+0.9}_{-7.4-12.3-0.9-0.6}$ \vspace{1ex} \\
\hline
\\$B^{0}_{s} \rightarrow \psi(3686)\overline{\emph{K}}^{*}_{2}(1430)^{0}(\rightarrow\emph{K}^-\pi^+)$  & $\cal{B}$ $(10^{-6})$  & $2.79^{+1.09+1.73+0.06+0.08}_{-0.80-1.30-0.03-0.07}$  & $2.98^{+1.03+1.70+0.05+0.07}_{-0.80-1.31-0.03-0.07}$  \vspace{1ex}  \\
& $f_{0}$ $(\%)$    &  $40.5^{+15.4+25.1+0.7+1.1}_{-11.5-18.3-0.4-1.1}$  & $36.9^{+15.1+23.8+0.7+1.0}_{-11.1-17.8-0.3-1.0}$ \vspace{1ex} \\
& $f_{\|}$ $(\%)$    &  $33.0^{+12.9+20.1+0.7+1.1}_{-9.3-15.4-0.4-0.7}$  & $33.6^{+11.1+18.1+0.7+1.0}_{-8.4-13.8-0.3-0.7}$ \vspace{1ex} \\
& $f_{\bot}$ $(\%)$    &  $26.5^{+10.8+16.8+0.7+0.7}_{-7.9-12.9-0.4-0.7}$  & $29.5^{+8.4+15.1+0.3+0.3}_{-7.4-12.4-0.3-0.7}$ \vspace{1ex} \\
\hline
\hline
\end{tabular*}
\end{table}

\begin{table}[htbp]
\centering
\caption{Branching ratios of the quasi-two-body decays $B^{0}_{s} \rightarrow \psi(3770)\overline{\emph{K}}^{*0}(\rightarrow\emph{K}^{-}\pi^{+})$ under the 2S-1D mixing mechanism calculated using the pQCD factorization approach. The first three uncertainties have been derived from the uncertainties in the previous tables, whereas the last one has been derived from the mixing angle.}
\label{tab7}
\begin{tabular*}{\columnwidth}{@{\extracolsep{\fill}}cccc@{}}
\hline
\hline
$\textmd{Decay mode}$  & $ $    & $\theta=(-12\pm2)^{\circ}$  & $\theta=(27\pm2)^{\circ}$   \\
\hline
\\$B^{0}_{s} \rightarrow \psi(3770)\overline{\emph{K}}^{*}_{0}(1430)^{0}(\rightarrow\emph{K}^-\pi^+)$  & $\cal{B}$ $(10^{-7})$  & $26.46^{+10.50+1.21+0.53+2.11}_{-7.18-1.75-0.31-2.10}$  & $1.30^{+0.49+0.31+0.08+0.37}_{-0.28-0.16-0.05-0.22}$  \vspace{1ex}  \\
\hline
\\$B^{0}_{s} \rightarrow \psi(3770)\overline{\emph{K}}^{*}(892)^{0}(\rightarrow\emph{K}^-\pi^+)$  & $\cal{B}$ $(10^{-6})$  & $10.29^{+3.25+6.60+0.63+0.78}_{-2.45-4.17-0.26-0.79}$  & $3.11^{+2.22+3.16+0.36+0.41}_{-1.46-1.78-0.24-0.38}$  \vspace{1ex}  \\
& $f_{0}$ $(\%)$    &  $17.4^{+12.8+1.0+0.8+3.7}_{-8.7-0.6-0.4-3.7}$  & $38.9^{+12.9+20.6+1.9+7.7}_{-8.0-16.1-1.3-7.1}$ \vspace{1ex} \\
& $f_{\|}$ $(\%)$    &  $40.7^{+8.2+31.0+2.0+1.6}_{-6.9-18.8-1.7-1.7}$  & $38.9^{+34.1+45.0+5.8+3.2}_{-24.8-25.1-3.9-2.9}$ \vspace{1ex} \\
& $f_{\bot}$ $(\%)$    &  $41.9^{+10.6+32.2+3.3+2.3}_{-8.2-21.2-2.4-2.3}$  & $22.2^{+24.4+36.0+3.9+2.3}_{-14.1-16.1-2.6-2.3}$ \vspace{1ex} \\
\hline
\\$B^{0}_{s} \rightarrow \psi(3770)\overline{\emph{K}}^{*}(1410)^{0}(\rightarrow\emph{K}^-\pi^+)$  & $\cal{B}$ $(10^{-8})$  & $18.13^{+5.31+12.56+1.18+1.72}_{-4.13-7.34-0.85-1.65}$  & $4.70^{+2.11+3.18+0.27+0.73}_{-1.55-1.67-0.21-0.60}$  \vspace{1ex}  \\
& $f_{0}$ $(\%)$    &  $19.8^{+9.2+1.3+2.9+3.8}_{-6.9-0.1-2.6-3.6}$  & $38.5^{+10.9+18.1+1.1+9.6}_{-8.7-16.0-0.6-8.5}$ \vspace{1ex} \\
& $f_{\|}$ $(\%)$    &  $40.4^{+9.2+28.6+1.0+2.4}_{-7.4-19.8-0.7-2.3}$  & $39.6^{+23.8+32.6+1.5+3.0}_{-17.2-12.6-1.1-2.3}$ \vspace{1ex} \\
& $f_{\bot}$ $(\%)$    &  $39.8^{+10.9+35.4+2.6+3.3}_{-8.5-20.6-1.4-3.2}$  & $21.9^{+15.4+17.0+3.2+3.0}_{-7.0-6.8-2.8-1.9}$ \vspace{1ex} \\
\hline
\\$B^{0}_{s} \rightarrow \psi(3770)\overline{\emph{K}}^{*}(1680)^{0}(\rightarrow\emph{K}^-\pi^+)$  & $\cal{B}$ $(10^{-8})$  & $6.53^{+2.04+4.69+0.32+0.66}_{-1.57-2.85-0.22-0.62}$  & $1.73^{+0.85+1.29+0.14+0.29}_{-0.61-0.65-0.08-0.23}$  \vspace{1ex}  \\
& $f_{0}$ $(\%)$    &  $19.9^{+10.1+1.1+2.5+3.8}_{-7.5-0.5-1.7-3.8}$  & $38.2^{+11.6+20.8+1.7+10.4}_{-8.7-16.2-0.6-8.7}$ \vspace{1ex} \\
& $f_{\|}$ $(\%)$    &  $40.3^{+9.6+34.5+1.1+2.6}_{-7.7-21.0-0.6-2.5}$  & $40.0^{+26.0+34.7+2.3+3.7}_{-18.5-13.9-1.2-2.3}$ \vspace{1ex} \\
& $f_{\bot}$ $(\%)$    &  $39.8^{+11.5+36.3+1.4+3.5}_{-8.9-22.2-1.1-3.2}$  & $21.8^{+11.6+19.1+4.0+2.9}_{-8.1-7.5-2.9-2.3}$ \vspace{1ex} \\
\hline
\\$B^{0}_{s} \rightarrow \psi(3770)(\emph{K}^-\pi^+)_{\emph{P}}$  & $\cal{B}$ $(10^{-6})$  & $10.84^{+3.29+6.64+0.66+0.80}_{-2.48-4.20-0.49-0.78}$  & $3.38^{+2.25+3.19+0.36+0.42}_{-1.47-1.77-0.24-0.38}$  \vspace{1ex}  \\
& $f_{0}$ $(\%)$    &  $17.6^{+12.4+1.0+0.9+3.6}_{-8.3-0.5-0.5-3.4}$  & $38.5^{+12.1+19.2+1.8+7.1}_{-8.3-14.8-1.2-6.5}$ \vspace{1ex} \\
& $f_{\|}$ $(\%)$    &  $40.6^{+7.8+29.5+1.9+1.6}_{-6.9-18.4-1.6-1.6}$  & $39.1^{+31.7+41.7+5.3+3.0}_{-22.5-23.1-3.6-2.7}$ \vspace{1ex} \\
& $f_{\bot}$ $(\%)$    &  $41.8^{+10.1+30.7+3.2+2.2}_{-7.7-19.9-2.5-2.2}$  & $22.4^{+22.8+33.4+3.6+2.4}_{-12.7-14.5-2.4-2.1}$ \vspace{1ex} \\
\hline
\\$B^{0}_{s} \rightarrow \psi(3770)\overline{\emph{K}}^{*}_{2}(1430)^{0}(\rightarrow\emph{K}^-\pi^+)$  & $\cal{B}$ $(10^{-7})$  & $5.40^{+1.44+2.52+0.12+0.76}_{-1.19-1.95-0.09-0.69}$  & $3.29^{+1.17+1.73+0.11+0.74}_{-0.84-1.30-0.07-0.69}$  \vspace{1ex}  \\
& $f_{0}$ $(\%)$    &  $25.7^{+12.8+18.1+0.9+5.9}_{-9.1-14.3-0.7-5.4}$  & $45.3^{+14.0+25.5+1.2+8.8}_{-10.9-18.5-0.6-8.2}$ \vspace{1ex} \\
& $f_{\|}$ $(\%)$    &  $32.8^{+7.0+13.5+0.7+4.4}_{-5.7-8.7-0.6-4.1}$  & $29.8^{+9.7+13.1+1.2+7.6}_{-7.3-10.9-0.9-7.3}$ \vspace{1ex} \\
& $f_{\bot}$ $(\%)$    &  $41.5^{+6.9+15.0+0.6+3.7}_{-7.2-13.1-0.4-3.3}$  & $24.9^{+11.9+14.0+0.9+6.1}_{-7.3-10.0-0.6-5.5}$ \vspace{1ex} \\
\hline
\hline
\end{tabular*}
\end{table}

Considering the Clebsch-Gorden coefficients, we can write the following relation
\begin{equation}
\begin{split}
\bigg|\emph{K}\pi,\emph{I}=\frac{1}{2}\bigg{\rangle}=\sqrt{\frac{1}{3}}|\overline{\emph{K}}^{0}\pi^{0}\rangle-\sqrt{\frac{2}{3}}|\emph{K}^{-}\pi^{+}\rangle.
\end{split}
\end{equation}

In our calculation, for the quasi-two-body decay $B^{0}_{s} \rightarrow \psi\overline{\emph{K}}^{*0}\rightarrow\psi\emph{K}^-\pi^+$, isospin conservation was assumed for the strong decays of an $\emph{I}=1/2$ intermediate resonance $\overline{\emph{K}}^{*0}$ to $\emph{K} \pi$, which can be expressed as follows:
\begin{equation}
\begin{split}
\frac{\Gamma(\overline{\emph{K}}^{*0}\rightarrow \overline{\emph{K}}^{0}\pi^{0})}{\Gamma(\overline{\emph{K}}^{*0}\rightarrow \emph{K}\pi)}=\frac{1}{3},
\frac{\Gamma(\overline{\emph{K}}^{*0}\rightarrow \emph{K}^{-}\pi^{+})}{\Gamma(\overline{\emph{K}}^{*0}\rightarrow \emph{K}\pi)}=\frac{2}{3}.
\end{split}
\end{equation}

Therefore, the branching ratios of $B^{0}_{s} \rightarrow \psi(2S,1D)\overline{\emph{K}}^{*0}(\rightarrow \emph{K} \pi)$ and $B^{0}_{s} \rightarrow \psi(3686,3770)\overline{\emph{K}}^{*0}(\rightarrow \emph{K} \pi)$ decays can be extracted directly under the narrow-width approximation relation
\begin{equation}
{\cal B}(B^{0}_{s}\rightarrow \psi \overline{\emph{K}}^{*0} \rightarrow \psi \emph{K}^{-}\pi^{+})=
{\cal B}(B^{0}_{s}\rightarrow \psi \overline{\emph{K}}^{*0}) \cdot {\cal B}(\overline{\emph{K}}^{*0}\rightarrow \emph{K}\pi)\cdot \frac{2}{3}.
\end{equation}

A comparison of the branching ratios for $\psi(3770)$ decay modes when the mixing angle is set to $\theta=-12^{\circ}$ and $\theta=27^{\circ}$ reveal a significant difference between the two choices, which can be attributed to the visibly small decay constant of $\psi(1D)$ compared to that of $\psi(2S)$. These results are in accordance with the analyses presented in other studies ~\cite{4435621991,283612006,770540032008,1730901978,2921361984}. In addition, when the 2S-1D mixing scheme is considered for the $B^{0}_{s}\rightarrow \psi(3686)\emph{K}^-\pi^+$ decay, the numerical result changes slightly compared to that of the $B^{0}_{s}\rightarrow \psi(2S)\emph{K}^-\pi^+$ decay, indicating that the $\psi(3686)$ state might be deemed as the $\psi(2S)$ state. Further, according the Eqs.~(\ref{eq:exp40}) and ~(\ref{eq:exp41}), the reason for the $\psi(3686)$ and $\psi(3770)$ decay modes having markedly different sensitivities to the change in the mixing angle under the 2S-1D mixing scheme could be provided. Numerically, ${\cal A}(B^{0}_{s}\rightarrow \psi(2S)\emph{K}^{-}\pi^+)$ is much larger than ${\cal A}(B^{0}_{s}\rightarrow \psi(1D)\emph{K}^{-}\pi^+)$, and thus the former dominates the decay amplitudes of the $\psi(3686)$ as well as $\psi(3770)$ decay modes. The value of the amplitude $\sin\theta {\cal A}(B^{0}_{s}\rightarrow \psi(2S)\emph{K}^{-}\pi^+)$ is greatly changed when the mixing angle is switched between $\theta=-12^{\circ}$ and $\theta=27^{\circ}$. On the contrary, the amplitude $\cos\theta {\cal A}(B^{0}_{s}\rightarrow \psi(2S)\emph{K}^{-}\pi^+)$ is relatively stable under this switch. Thus, the branching ratio of the decay $B^{0}_{s} \rightarrow \psi(3686)\overline{\emph{K}}^{*0}(\rightarrow\emph{K}^{-}\pi^{+})$ is stable under the switch between the two values of the mixing angle, whereas the branching ratio of the decay $B^{0}_{s} \rightarrow \psi(3770)\overline{\emph{K}}^{*0}(\rightarrow\emph{K}^{-}\pi^{+})$ is highly sensitive to the variation of the mixing angle. The running LHCb experiment is an excellent place to detect decays $B^{0}_{s} \rightarrow \psi(3686,3770)\overline{\emph{K}}^{*0}(\rightarrow\emph{K}^-\pi^+)$ with branching ratios of the order of $10^{-5}-10^{-8}$, which will help us gain a better understanding about the mixing mechanism of the charmonium mesons.

\section{Summary} \label{sec:summary}

In this work, we have studied the $B^{0}_{s}\rightarrow \psi(2S,1D)\overline{\emph{K}}^{*0}(\rightarrow\emph{K}^-\pi^+)$ decays using the pQCD factorization approach by introducing the kaon-pion DAs. We considered the S-wave resonance $\overline{\emph{K}}^{*}_{0}(1430)^{0}$, the P-wave resonances $\overline{\emph{K}}^{*}(892)^{0}$, $\overline{\emph{K}}^{*}(1410)^{0}$, and $\overline{\emph{K}}^{*}(1680)^{0}$, and the D-wave resonance $\overline{\emph{K}}^{*}_{2}(1430)^{0}$. This study covers three types of polarization amplitudes, namely, longitudinal, parallel, and vertical, which reflect the role of the different polarization conditions in the decay in terms of the polarization fractions. Based on the 2S-1D mixing scheme, we have obtained the branching ratios of the $B^{0}_{s} \rightarrow \psi(3686,3770)\emph{K}^{-}\pi^{+}$  decays by fitting the decay amplitudes of $\psi(2S)$ and $\psi(1D)$ decay modes. Finally, the pQCD predictions for the $B^{0}_{s} \rightarrow \psi(2S,1D)\emph{K} \pi$ and $B^{0}_{s} \rightarrow \psi(3686,3770)\emph{K} \pi$ decays have been obtained using a narrow-width approximation relation.

The pQCD predictions indicate that the $\overline{\emph{K}}^{*}(892)^{0}$ resonance is the main contributor to the total decay, and the branching ratios of the $\psi(2S)$ decay modes agree well with the existing experimental data within acceptable errors. Our calculations show that the branching ratios of the $\psi(3686)$ and $\psi(2S)$ decay modes are very similar, suggesting that they can be regarded as the same state. Theoretical predictions for the branching ratios of $\psi(3686)$ and $\psi(3770)$ decay channels are of the order of $10^{-5}$ and $10^{-6}$, respectively, which will be verified using the data from future experimental measurements. The detected data will help us to gain further understanding about the internal structures of the $\psi(3686)$ and $\psi(3770)$ mesons.


\section*{acknowledgments}

This work has been supported by the National Natural Science Foundation of China under Grant No.11047028 and by the Fundamental Research Funds of the Central Universities, Grant Number XDJK2012C040.

\section*{Appendix : functions involved in the calculation} \label{sec:appendix}
\appendix

Important formulae used in the calculations are listed in this section. The Sudakov exponents in the decay amplitudes are defined as
\renewcommand{\theequation}{{A}.1}
\begin{eqnarray}
\begin{split}
&S_{B^0_{s}}=s(x_{B}p^+_{1},b_{B})+\frac{5}{3}\int^{t}_{1/b_{B}}\emph{d}{\bar{\mu}}\frac{\gamma_{q}(\alpha_{s}({\bar{\mu}}))}{\bar{\mu}},\\
&S_{M}=s(\bar{z}p^+,b)+s(zp^{+},b)+2\int^{t}_{1/b}\emph{d}{\bar{\mu}}\frac{\gamma_{q}(\alpha_{s}({\bar{\mu}}))}{\bar{\mu}},\\
&S_{\psi}=s_{\emph{c}}(\bar{x}_{3}p^-_{3},b_{3})+s_{\emph{c}}({x}_{3}p^{-}_{3},b_{3})+2\int^{t}_{m_{c}}\emph{d}{\bar{\mu}}\frac{\gamma_{q}(\alpha_{s}({\bar{\mu}}))}{\bar{\mu}},
\end{split}
\end{eqnarray}
where the Sudakov factors, $s(Q,b)$ and $s_{\emph{c}}(Q,b)$, have been derived from the resummation of the double logarithms. Their specific expressions can be found in the Refs.~\cite{5551972003,971130012018}.

The parameterized expression of the threshold resummation function $S_{t}(x)$ is~\cite{650140072001}
\renewcommand{\theequation}{{A}.2}
\begin{equation}
S_{t}(x)=[x(1-x)]^{c}\frac{2^{1+2c}\Gamma(\frac{3}{2}+c)}{\sqrt{\pi}\Gamma(1+c)},
\end{equation}
where $c=0.04Q^2-0.51Q+1.87$ and $Q=\sqrt{M_{B_{s}^{0}}^{2}(1-r^2)}$~\cite{800740242009}.

The hard scattering kernel functions $h_{i}$ in the decay amplitudes have been derived from the Fourier transform of the virtual quark and the gluon propagators, which can be specifically expressed as
\renewcommand{\theequation}{{A}.3}
\begin{equation}
\begin{split}
&h_{a}(x_{B}, z, b_{B}, b)= K_{0}(M_{B_{s}^{0}}b_{B}\sqrt{(1-r^2)x_{B}z}) [\theta(b-b_{B})I_{0}(M_{B_{s}^{0}}b_{B}\sqrt{(1-r^2)z})K_{0}(M_{B_{s}^{0}}b\sqrt{(1-r^2)z})+(b_{B}\leftrightarrow b)],\\
&h_{b}(x_{B}, z, b_{B}, b)= K_{0}(M_{B_{s}^{0}}b\sqrt{(1-r^2)x_{B}z}) \\
&\quad \quad \quad \quad\quad\quad\quad\times \begin{cases}
[\frac{\emph{i}{\pi}}{2} \theta(b-b_{B})J_{0}(M_{B_{s}^{0}}b_{B}\sqrt{|\kappa|})H^{(1)}_{0}(M_{B_{s}^{0}}b\sqrt{|\kappa|})+(b_{B}\leftrightarrow b)], & {\kappa < 0}\\
[\theta(b-b_{B})I_{0}(M_{B_{s}^{0}}b_{B}\sqrt{\kappa})K_{0}(M_{B_{s}^{0}}b\sqrt{\kappa})+(b_{B}\leftrightarrow b)], & {\kappa \geq 0}\\
\end{cases}\\
&h_{c}(x_{B}, z, x_{3}, b_{B}, b_{3})=[\theta(b_{3}-b_{B})I_{0}(M_{B_{s}^{0}}b_{B}\sqrt{(1-r^2)x_{B}z})K_{0}(M_{B_{s}^{0}}b_{3}\sqrt{(1-r^2)x_{B}z})+(b_{B}\leftrightarrow b_{3})]  \\
 &\quad \quad \quad \quad\quad\quad\quad\quad\quad\times \begin{cases}
\frac{\emph{i}{\pi}}{2} H^{(1)}_{0}(M_{B_{s}^{0}}b_{3}\sqrt{|\beta|}), & {\beta < 0}\\
K_{0}(M_{B_{s}^{0}}b_{3}\sqrt{\beta}), & {\beta \geq 0}\\
\end{cases}\\
&h_{d}(x_{B}, z, x_{3}, b_{B}, b_{3})=h_{c}(x_{B}, z, \bar{x}_{3}, b_{B}, b_{3}),
\end{split}
\end{equation}
where $\kappa=(x_{B}-\eta)(1-r^2)$ and $\beta=r^2_{c}-(\bar{x}_{3}r^2+(1-r^2)z)(\bar{x}_{3}\bar{\eta}-x_{B})$. $I_{0}$ , $K_{0}$ are the modified Bessel functions and $J_{0}$ is the Bessel function with $H^{(1)}_{0}(x)=\emph{i}Y_{0}(x)+J_{0}(x)$.

To eliminate the radiative corrections of large logarithms, the hard scales, $t_{i}$, in the decay amplitudes are chosen as
\renewcommand{\theequation}{{A}.4}
\begin{equation}
\begin{split}
&t_{a}=\mathrm{Max}\{M_{B_{s}^{0}}\sqrt{(1-r^2)z}, \frac{1}{b_{B}}, \frac{1}{b}\},\\
&t_{b}=\mathrm{Max}\{M_{B_{s}^{0}}\sqrt{|\kappa|}, \frac{1}{b_{B}}, \frac{1}{b}\},\\
&t_{c}=\mathrm{Max}\{M_{B_{s}^{0}}\sqrt{(1-r^2)x_{B}z}, M_{B_{s}^{0}}\sqrt{|\beta|}, \frac{1}{b_{B}}, \frac{1}{b_{3}}\},\\
&t_{d}=t_{c}|_{x_{3} \rightarrow \bar{x}_{3}}.
\end{split}
\end{equation}


\end{document}